\documentclass[twocolumn]{aastex631}
\usepackage{graphicx}
\usepackage[caption=false]{subfig}
\usepackage{amsmath,amsfonts,amsthm,bm}
\usepackage{amssymb}
\usepackage{rotating}
\usepackage{tabularx}
\usepackage{comment}
\usepackage{xcolor}
\usepackage{rotating}
\usepackage{mathtools}
\usepackage{chngcntr}

\newcommand*{\WidestExpression}{(B+t_d)/P}
\newcommand*{\MakeBox}[1]{\makebox[\widthof{$\WidestExpression$}][l]{$#1$}}

\definecolor{new_color}{RGB}{189,39,7}
\newcommand\bedit[1]{{#1}}
\newcommand\beditr[1]{{#1}}%{\textcolor{new_color}{\textbf{#1}}}

\providecommand{\dedtNoPlanetC}{\ensuremath{1.75_{-1.68}^{+1.71}\cdot 10^{-3}}}
\providecommand{\domdtNoPlanetC}{\ensuremath{1.12_{-0.22}^{+0.22}}}

\providecommand{\dedtLinearPlanetC}{\ensuremath{3.28_{-1.75}^{+1.72} \cdot 10^{-3}}}
\providecommand{\domdtLinearPlanetC}{\ensuremath{1.12_{-0.22}^{+0.22}}}

\providecommand{\dedtQuadPlanetC}{\ensuremath{7.7_{-18.6}^{+18.6} \cdot 10^{-4}}}
\providecommand{\domdtQuadPlanetC}{\ensuremath{0.76_{-0.24}^{+0.24}}
}

\providecommand{\dedtKepPlanetC}{\ensuremath{-7.3_{-18.0}^{+19.0} \cdot 10^{-4}}}
\providecommand{\domdtKepPlanetC}{\ensuremath{0.228_{-0.11}^{+0.11}}}

\received{May 5, 2023}
\revised{July 28, 2023}
\accepted{August 3, 2023}
\published{September 1, 2023}

\shorttitle{AASTeX v6.3.1 Sample article}
\shortauthors{de Beurs et al.}
\graphicspath{{./}{figures/}}
\begin{document}

%\title{\beditr{Confirming a companion to} HAT-P-2 \beditr{b}  and \beditr{reconsidering its} rapid orbital evolution}

\title{\beditr{Revisiting Orbital Evolution in HAT-P-2 b and Confirmation of HAT-P-2 c}}

\vspace{2in}
\author[0000-0002-7564-6047]{Zo{\"e}\ L. de Beurs}
\affiliation{Department of Earth, Atmospheric and Planetary Sciences, Massachusetts Institute of Technology, Cambridge, MA 02139, USA}
\affiliation{NSF Graduate Research Fellow and MIT Presidential Fellow}

\author[0000-0003-2415-2191]{Julien de Wit}
\affiliation{Department of Earth, Atmospheric and Planetary Sciences, Massachusetts Institute of Technology, Cambridge, MA 02139, USA}

\author[0000-0002-8400-1646]{Alexander Venner}
\affiliation{Centre for Astrophysics, University of Southern Queensland, Toowoomba, QLD 4350, Australia}

\author[0000-0001-6298-412X]{David Berardo}
\affiliation{Department of Physics and Kavli Institute for Astrophysics and Space Research, Massachusetts Institute of Technology, Cambridge, MA 02139, USA}

\author[0000-0001-9466-1843]{Jared Bryan}
\affiliation{Department of Earth, Atmospheric and Planetary Sciences, Massachusetts Institute of Technology, Cambridge, MA 02139, USA}
\affiliation{NSF Graduate Research Fellow and MIT Presidential Fellow}

\author[0000-0002-4265-047X]{Joshua N. Winn}
\affiliation{Department of Astrophysical Sciences, Princeton University, Princeton, NJ 08544, USA}

\author[0000-0003-3504-5316]{Benjamin J. Fulton}
\affiliation{Cahill Center for Astronomy \& Astrophysics, California Institute of Technology, Pasadena, CA 91125, USA}

\author[0000-0001-8638-0320]{Andrew W. Howard}
\affiliation{Department of Astronomy, California Institute of Technology, Pasadena, CA 91125, USA}

%\author{Erik Petigura}
%\affiliation{Department of Physics \& Astronomy, University of California, Los Angeles, USA}

%\collaboration{6}{\textcolor{black}{California Planet Search Team}}

%% Note that the \and command from previous versions of AASTeX is now
%% depreciated in this version as it is no longer necessary. AASTeX 
%% automatically takes care of all commas and "and"s between authors names.

%% AASTeX 6.31 has the new \collaboration and \nocollaboration commands to
%% provide the collaboration status of a group of authors. These commands 
%% can be used either before or after the list of corresponding authors. The
%% argument for \collaboration is the collaboration identifier. Authors are
%% encouraged to surround collaboration identifiers with ()s. The 
%% \nocollaboration command takes no argument and exists to indicate that
%% the nearby authors are not part of surrounding collaborations.

%% Mark off the abstract in the ``abstract'' environment. 
\begin{abstract}
One possible formation mechanism for Hot Jupiters is that high-eccentricity gas giants experience tidal interactions with their host star that cause them to lose orbital energy and migrate inwards. We study these types of tidal interactions in an eccentric Hot Jupiter called HAT-P-2 b, which is a system where a long-period companion has been suggested, and hints of orbital evolution \citep{deWit2017} were detected. Using five additional years of radial velocity (RV) measurements, we further investigate these phenomena. We investigated the long-period companion by jointly fitting RVs and \textit{Hipparcos-Gaia} astrometry and confirmed this long-period companion, significantly narrowed down the range of possible periods ($P_2 = 8500_{-1500}^{+2600}$ days), and determined that it must be a substellar object ($10.7_{-2.2}^{+5.2}$ $M_j$). We also developed a modular pipeline to simultaneously model rapid orbital evolution and the long-period companion. We find that the rate and significance of evolution are highly dependent on the long-period companion modeling choices. In some cases the orbital rates of change reached $de/dt = {3.28}_{-1.72}^{+1.75} \cdot \textcolor{black}{10^{-3}}$/year, $d\omega/dt = 1.12 \pm 0.22 ^{\circ}$/year which corresponds to a $\sim 321$ year apsidal precession period. In other cases, the data is consistent with $de/dt = 7.67 \pm 18.6  \textcolor{black}{\cdot 10^{-4}}$/year, $d\omega/dt = 0.76\pm 0.24 ^{\circ}$/year. The most rapid changes found are significantly larger than the expected relativistic precession rate and could be caused by transient tidal planet-star interactions.  To definitively determine the magnitude and significance of potential orbital evolution in HAT-P-2 b, we recommend further monitoring with RVs and precise transit and eclipse timings.
\end{abstract}


%% From the front matter, we move on to the body of the paper.
%% Sections are demarcated by \section and \subsection, respectively.
%% Observe the use of the LaTeX \label
%% command after the \subsection to give a symbolic KEY to the
%% subsection for cross-referencing in a \ref command.
%% You can use LaTeX's \ref and \label commands to keep track of
%% cross-references to sections, equations, tables, and figures.
%% That way, if you change the order of any elements, LaTeX will
%% automatically renumber them.
%%
%% We recommend that authors also use the natbib \citep
%% and \citet commands to identify citations.  The citations are
%% tied to the reference list via symbolic KEYs. The KEY corresponds
%% to the KEY in the \bibitem in the reference list below. 

\section{Introduction}\label{sec:intro}
The role of planet-star interactions \bedit{in} the migration of Hot Jupiters is not well understood. The original discovery of Jupiter-sized planets \bedit{on} short-period orbits \citep{mayor1995} challenged our theories of solar system formation and it was proposed that Hot Jupiters  must have migrated inward after forming beyond the ice line \citep{1996Natur.380..606L}. One possible migration mechanism proposed is that high-eccentricity gas giants experience tidal interactions with their host star that cause them to lose orbital energy and migrate to a close-in orbit  \bedit{\citep{Rasio_Ford_1996, Weidenschilling_Marzari_1996, Lin_Ida_1997}}. %These tidal interactions may leave both observational signatures in photometric data and radial velocity measurements for high-eccentricity Hot Jupiter systems.

High-eccentricity tidal migration theories often focus on the tides within the planet. However, equilibrium tides \citep{1981A&A....99..126H, 1998ApJ...499..853E} in the star are accompanied by dynamical tides \citep{1998ApJ...507..938G, Fuller2012} \beditr{consisting of resonantly excited g-modes}, which may result in energy and angular momentum exchanges between the star and orbit that can be several orders of magnitude larger \citep{Witte2002, Ma2021}. The resulting changes in the orbit could impact radial velocity and photometric observations of a planetary system.  Hints of rapid orbital evolution were detected in the HAT-P-2 system where other signs of strong planet-star interactions have been observed  \citep{deWit2017}.

HAT-P-2 is an F8 type star and hosts an $8.70_{-0.\bedit{20}}^{+0.\bedit{19}}$ $M_j$ mass planet on an eccentric ($e =0.51023 \pm 0.00042$), short period ($P = 5.6334675 \pm 0.0000013$ days) orbit. Due to this combination of characteristics, it has been identified as an ideal target for studying planet-star interactions \citep{2008Andres, Fabrycky2009, Levrard2009, Cowan2011, Lewis2013, Lewis2014, Salz2016}.

%In photometric observations, planet-star interactions could result in a specific type of stellar pulsations called tidally-excited oscillations (TEOs), which occur when the motion of an eccentric companion excites the natural oscillation modes of a star \citep{1995ApJ...449..294K}. Rather than oscillating at the frequency of a star's natural stellar modes, TEOs instead oscillate at integer multiples of the system's orbital frequency \citep{2020svos.conf..203G, 2017MNRAS.472.1538F, 2012MNRAS.420.3126F}. The amplitudes of these signals equal the difference between the frequency of a natural mode and the tidal forcing frequency \citep{2018MNRAS.473.5165H, 1995ApJ...449..294K} and can thus result in a unique signal in photometric data. 

%These kind of tidally induced oscillations had originally only been observed in highly eccentric stellar binaries. However, in 2017, \citet{deWit2017} discovered the first system where a planet induced tidal pulsations upon its host star: HAT-P-2b. In this system, the pulsation modes match the exact harmonics of the planet's orbital frequency. 

Now, we are revisiting this interesting system because the California Planet Search (CPS) team observed HAT-P-2 for an additional five years with the HIRES spectrograph, which extends the RV baseline from nine years to fourteen years. In addition, we also merged RVs from multiple instruments and increased the number of RVs from 54 to 103 compared to \citet{deWit2017}. With these additional RV measurements, we were able to further investigate whether the hints of rapid orbital evolution appear to persist and can be confirmed. We developed our own modular orbital evolution pipeline that allows for an apples-to-apples comparison between various orbital models. 

In addition, we performed an in-depth asses\bedit{s}ment of the possibility of an outer long-period companion as has been suggested by \citet{Lewis2013, Knutson2014, 2017Bonomo, Ment2018}. We combined our RV measurements with astrometric observations and confirmed a persistent signal that caused by a substellar long-period companion HAT-P-2 c. 

Our paper is organized as follows. In Section 2, we describe the RV and astrometric data included in this analysis, and in Section 3, we describe RV orbital fitting pipeline and how we performed a joint astrometric-RV fit. In Section 4, we describe our results and to evalulate our models. In Section 5, we describe the interpretation of these results. Lastly, in Section 6 and 7, we provide directions for future work and \beditr{we} conclude.

%\bpurple{My initial results are promising; I was able to confirm a rapid change in argument of periastron ($\omega$) and eccentricity ($e$), corresponding to a $7.2\%$ increase and $3.7\%$ decrease respectively over the course of 14 years.} This change in $\omega$ is significantly larger than what would be expected from general relativity alone and could potentially be explained by tidal planet-star interactions. Currently, I am working to perform a simultaneous fit with the RV and transit observations and hope to model our observations with stellar oscillation and evolution codes.

 %and 10 new transits of HAT-P-2b have also been observed with the Transiting Exoplanet Survey Satellite \citep[TESS;][]{2015JATIS...1a4003R}.

\section{Data}

\subsection{Precise Radial Velocity Measurements}

Our analysis includes 72 spectra of HAT-P-2 obtained by the California Planet Search (CPS) Team using the High Resolution Echelle Spectrograph (HIRES) on the W.M. Keck Observatory 10m telescope Keck-I \citep{1994Vogt}. The standard CPS data reduction pipeline was used, which uses an iodine cell as a wavelength reference as further described in \citet{2010ApJ...721.1467H}. These \bedit{72} RVs span 14 years (2006-09-03 through 2020-08-31). \bedit{For comparison, \citep{deWit2017} used 44 HIRES RVs that span 9 years (2006-09 through 2015-10).}

In addition \bedit{to our 72 HIRES RVs}, we \bedit{also} included 31 publicly available RVs from the HARPS-N spectrograph at the 3.6m Telescopio Nazionale Galileo on La Palma in the Canary Islands. HARPS-N is a temperature and pressure stabilized echelle spectrograph \citep{2012SPIE.8446E..1VC}. HARPS-N spans the wavelength range from 383 to 693 nm and has a resolving power of $\lambda/\Delta\lambda=$ 115,000. These RVs span 2013-03-12 through 2017-09-28. We included the published RVs from \citet{2017Bonomo} and downloaded additional publicly available RVs through the TNG Archive\footnote{\hyperref[http://archives.ia2.inaf.it/tng/]{http://archives.ia2.inaf.it/tng/}}. 

We removed RV measurements that occur during transit \citep{Winn2007b} such that we are not using RVs that are affected by the Rossiter–McLaughlin effect. \beditr{The RVs are included in the appendix in Table \ref{harpsntable}.}

\subsection{Astrometric Measurements}
The \bedit{astrometric data} used in this work are derived from the \textit{Hipparcos-Gaia} Catalogue of Accelerations \citep[HGCA;][]{2018Brandt, Brandt2021}. \bedit{This} consists of three independent proper motion measurements: the \textit{Hipparcos} proper motion ($\mu_H$), the \textit{Gaia} DR3 proper motion ($\mu_G$\bedit{)}, and the mean \textit{Hipparcos-Gaia} proper motion ($\mu_{HG}$). \bedit{See \citet{2018Brandt} and  \citet{Brandt2019} for further details. The astrometric data has a time baseline of approximately 25 years (1991-2016)}.

RVs provide us with the minimum mass $M_p \sin(i)$ of an object, where $i$ is the inclination of the system. However, when combining RVs with astrometric measurements, we can constrain the inclination of the system and derive the actual mass ($M_p$). In this way, astrometric measurements can help constrain the nature of \bedit{the outer companion of HAT-P-2} (stellar, brown dwarf, planet).

\section{Data Analysis}

\subsection{Radial Velocity Orbital fitting pipeline}\label{pipeline_descript}
We designed a modular RV orbital fitting pipeline that allows for comparison of various dynamical scenarios to determine the origins of the RV variation seen in \citet{deWit2017}. We use  \bedit{\texttt{radvel}'s \texttt{kepler.rv\_drive} function} \citep{RadVel_fulton_2018} to solve Kepler's equation in our pipeline and build upon this software by allowing orbital parameters to change over time. Our modular pipeline is parallelized to speed up orbital fitting and can easily be applied to other RV datasets. Our pipeline can notably perform fits of the following \beditr{stable} orbital configurations: (i) a stable, one-planet Keplerian fit, (ii) a stable, two-planet Keplerian fit, (iii) a stable, two-planet fit where the long-period companion is modeled as a quadratic trend. \beditr{In addition, our pipeline can model the eccentricity ($e$) and argument of periastron ($\omega$) as linearly evolving parameters such that $e = de/dt \cdot time + e_0$ and $\omega = d\omega/dt \cdot time + \omega_0$. We refer to models where $e$ and $\omega$ can linearly evolve with time as first-order, evolving models. In particular, our pipeline can perform the following evolving orbital configurations: } (iv) a first-order evolving, one-planet Keplerian fit, (v) a first-order evolving, two-planet Keplerian fit, (vi) and a first-order evolving, two-planet fit where the long-period companion is modeled as a quadratic trend.

In case (i), the model can be described by 13 free parameters, 5 of which describe the orbit (period ($P$), planet mass ($M_p$), time of periastrion passage ($t_p$),  eccentricity ($e$) and argument of periastron ($\omega$) which are parameterized as $\sqrt{e}\sin{\omega}$ and $\sqrt{e}\cos{\omega}$), 2 of which describe a linear  \bedit{($\dot{\gamma}$)} and quadratic \bedit{($\ddot{\gamma}$)} trend with time\footnote{In the case of HAT-P-2, we have a long-period secondary companion that has a period close to the baseline. For this reason, we do not include the linear an quadratic background trends to prevent degeneracies. Thus, we only have 11 free parameters for model (i) rather than 13.} and 6 which correspond to the offsets \bedit{($\gamma$)} and \beditr{RV} jitter terms for each of the RV datasets. \beditr{RV jitter describes the instrumental and astrophysical stochastic signals in the data. Astrophysical signals originate from stellar variability such as inhomogeneties on the stellar surface, pressure-mode oscillations, and granulation. Since each instrument has its own noise characteristics, it is common to fit for a jitter term for each instrument. }\bedit{We note that RV models are commonly parametrized using semi-amplitude (K) rather than mass. However, when $e$ and $\omega$ are allowed to vary over time for an evolving planet model, this parametrization allows mass to vary over time, which is unphysical. Parametrizing the RV model in terms of mass prevents this and instead only allows K to vary as a function of $e$ and $\omega$.} In case (ii), we gain an additional 5 orbital parameters for a secondary companion, which sums to 18 free parameters\footnote{We again do not include the linear and quadtratic trends for HAT-P-2 so we have 16 free parameters.}. In case (iii), the secondary companion is modeled as a second-order background trend so this adds a linear and quadratic term and thereby two additional free parameters. In case (iv), we add first-order polynomials for $e$ and/or $\omega$ and this adds one free parameter per polynomial. In case (v), we add an additional five parameters to fit for the orbital parameters of a secondary companion. Lastly, in case (vi), we instead model the secondary companion as a linear and quadratic trend so this adds two free parameters.

%In the first order evolving orbit scenario, we allow $e$ and/or $\omega$ to vary linearly with time by describing the eccentricity as $e = \frac{de}{dt}\cdot (time) + e_0$ and/or describing $\omega$ as $\omega = \frac{d\omega}{dt}\cdot (time) + \omega_0$. We then add $\frac{de}{dt}$ and/or $\frac{d\omega}{dt}$ as additional free parameters. 

%In the second order evolving orbit scenario, we also add a second order term to describe $e$ and/or $\omega$ such that 
%\begin{equation}
%    e =e_0  +\frac{de}{dt}\cdot (time) + %frac{de^2}{dt^2}\cdot (time)^2
%\end{equation}
%and such that 
%\begin{equation}
%    \omega = \omega_0  +\frac{d\omega}{dt}\cdot (time) + \frac{d \omega^2}{dt^2}\cdot (time)^2
%\end{equation}
%where $e_0$ and $\omega_0$ are the initial values of $e$ and $\omega$ respectively.These two additional free parameters result in a total of 24 free parameters.

Once a given orbital configuration is chosen, we sample our parameter space using \texttt{edmcmc\footnote{https://github.com/avanderburg/edmcmc}}, which is an implementation of Markov Chain Monte Carlo (MCMC) that incorporates differential evolution \citep[DE;][]{2006TerBraak}. DE is a genetic algorithm that solves the problem of choosi\bedit{n}g an appropriate scale and orientation for the jumping distribution, which significantly speeds up the fitting process. 

After performing the fit, the pipeline produces several diagnostic plots and performance metrics (described in Section \ref{perform_met}) to assess goodness of fit. Convergence of the parameters is determined using the Gelman-Rubin statistic \citep{1992Gelman} \bedit{where our Gelman-Rubin threshold $< 1.02$}. We computed \bedit{$10,000$} MCMC chains and discarded the first $25 \%$ burn-in samples to produce our posterior samples. We implemented Gaussian priors to the $P$ and  $t_p$ parameters based on transit and occultation times from previous studies \beditr{\citep{Ivshina2022}}. \bedit{For our models that include a Keplerian long-period companion, we also apply Gaussian priors to the companion's period ($P_2$) and time of periastron passage ($t_{p2}$) based on the astrometry analysis described in Section \ref{joint_astrometry_rv_fit} since the RVs do not sufficiently span its orbital period to constrain these parameters alone.} To derive our final parameter values, we computed the median and 68.3\% confidence intervals. The results for our comparison of models (i)-(vi) are described in Section \ref{Results}.

\begin{figure*}[p!]
\centering
	\includegraphics[width=0.85\textwidth]{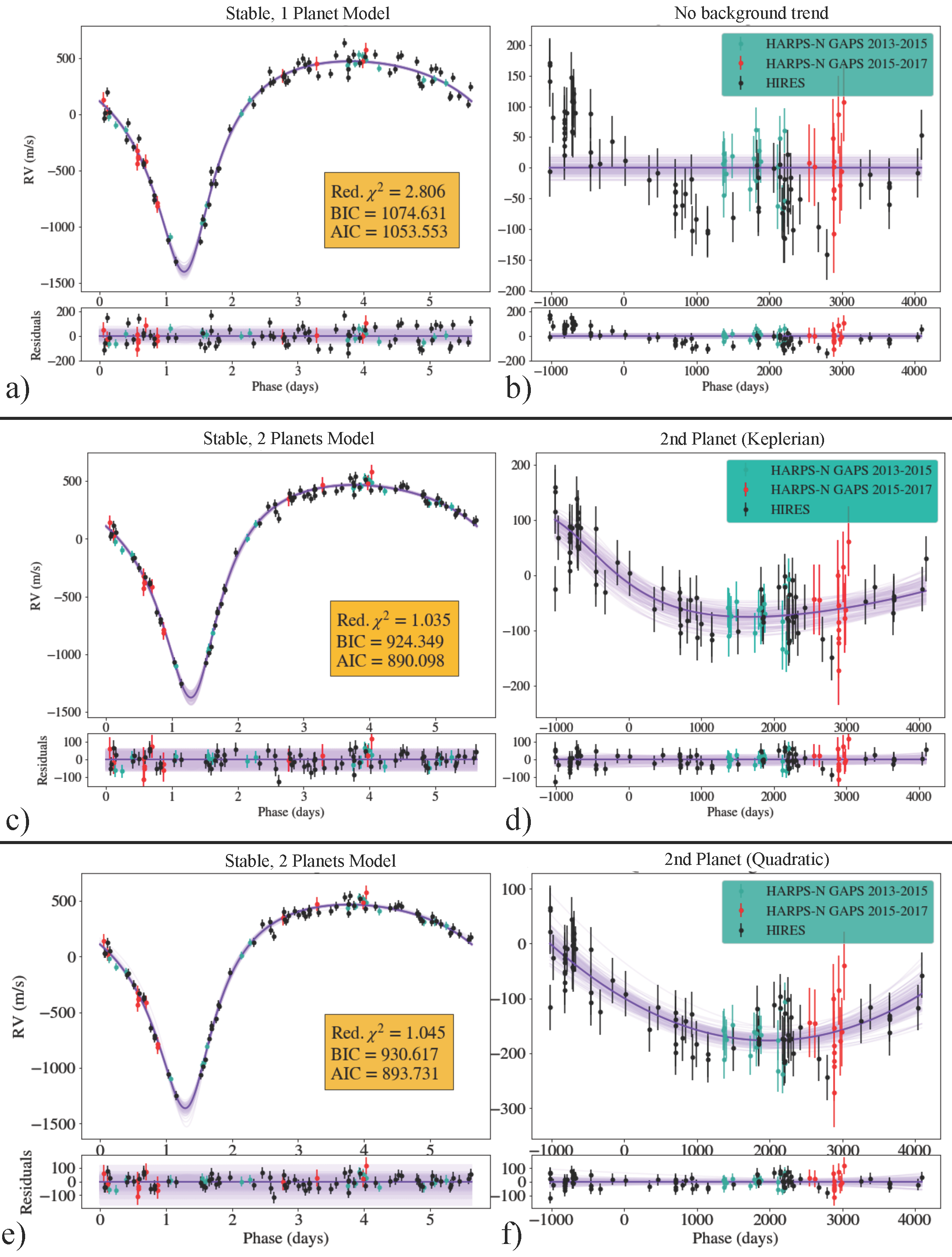}
    \caption{Comparison of our \bedit{stable, non-evolving} orbital models. From top to bottom, we include the stable one-planet model \bedit{with no background trends} (a, b), the stable two-planet model \bedit{with the long-period companion modeled as a Keplerian} (c,d), the \bedit{stable} two-planet model \bedit{with the long-period companion modeled as a quadratic trend}  (e,f). For more details on these orbital models, see Section \ref{pipeline_descript}. \bedit{In} each left hand panel (a, c, e), we plot the orbital fit of HAT-P-2 b minus a secondary companion (or no companion if this was not included in the model). The data is color-coded by instrument (\bedit{turquoise} = HARPS-N 2013-2015, red = HARPS-N 2015-2017, black = HIRES). The median model is plotted in dark purple and a 100 draws from the posteriors are plotted in light purple.  For each right hand panel (b, d, f), we plot the secondary companion (or no trend/companion if this was not included in the model) minus the median HAT-P-2 b model. \bedit{We also include the goodness-of-fit metrics corresponding to each orbital model in the yellow box. Red. $\chi_{\bedit{r}}^2$ is an abbreviation for reduced $\chi_{\bedit{r}}^2$.}}
    \label{fig:all_models}
\end{figure*}

\begin{figure*}[ht!]
\centering
	\includegraphics[width=0.85\textwidth]{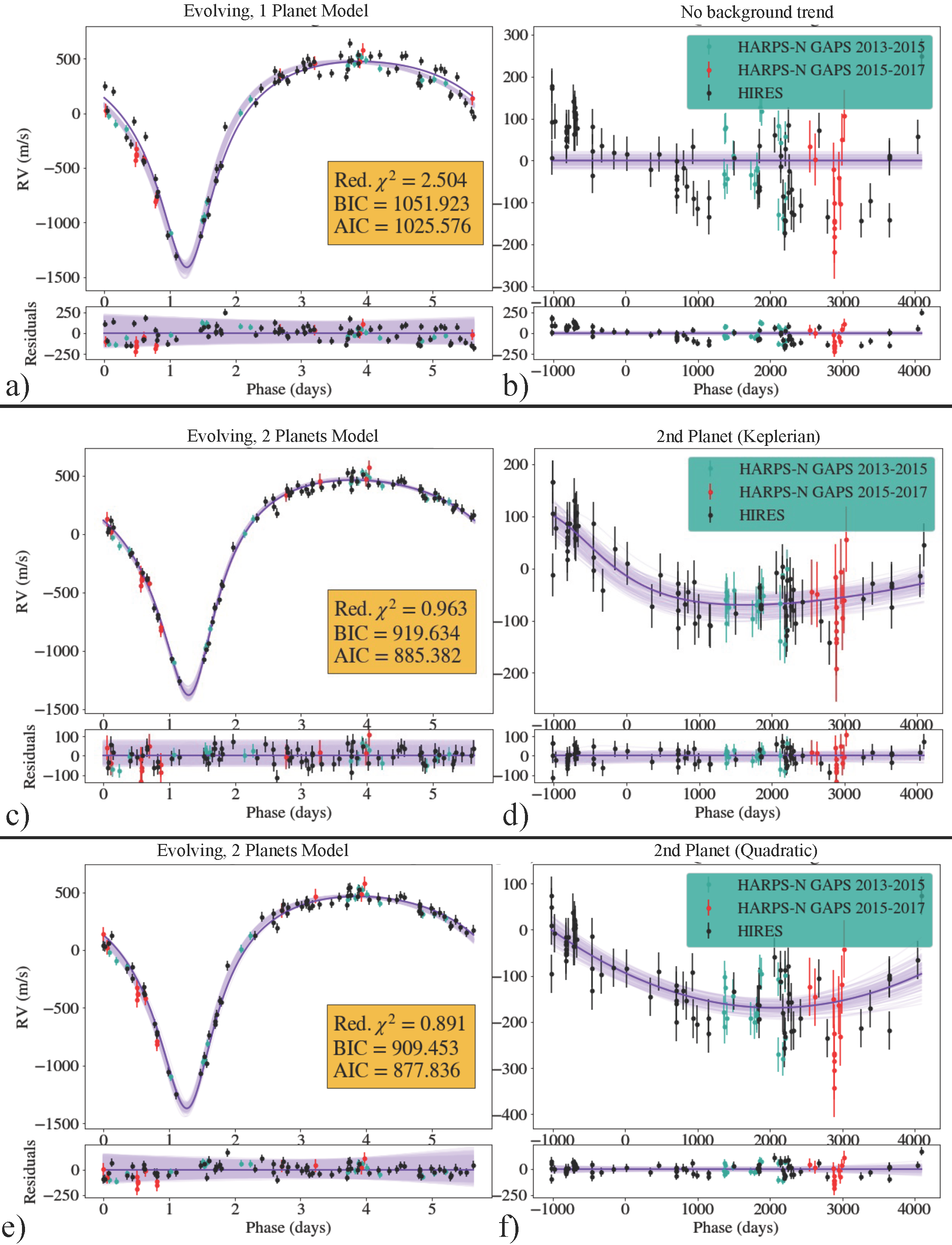}
    \caption{Comparison of our \bedit{Evolving} orbital models. From top to bottom, we include the evolving, one-planet model \bedit{with no background trends} (a, b), the evolving two-planet model \bedit{with the long-period companion modeled as a Keplerian} (c,d), the evolving two-planet model \bedit{with the long-period companion modeled as a quadratic trend} (e,f). For more details on these orbital models, see Section \ref{pipeline_descript}. The color-coding is identical to Figure \ref{fig:all_models}.}
    \label{fig:all_evolve_models}
\end{figure*}

% OG Fig Caption
%Comparison of our orbital models. From top to bottom, we include the stable one planet model (a, b), the stable two planet model (c,d), the evolving one planet model (e,f), and the evolving two planet model (g,h). For more details on these orbital models, see Section \ref{pipeline_descript}. For each left hand panel (a, c, e, g), we plot the orbital fit of HAT-P-2 b minus a background trend or secondary companion and include the goodness-of-fit statistics (reduced $\chi_{\bedit{r}}^2$, BIC, AIC). The data is color-coded by instrument (\bedit{turquoise} = HARPS-N 2013-2015, red = HARPS-N 2015-2017, black = HIRES). The median model is plotted in dark purple and a 100 draws from the posteriors are plotted in light purple.  For each right hand panel (b, d, f, h), we plot the background trend or secondary companion (or no trend if this was not included in the model) minus the median HAT-P-2 b model. \bedit{We also include the goodness-of-fit metrics corresponding to each orbital model in the yellow box. Red. $\chi_{\bedit{r}}^2$ is an abbreviation for reduced $\chi_{\bedit{r}}^2$.}

\subsection{Performance metrics}\label{perform_met}
We use four different metrics to assess the goodness of fit of our various orbital models: (1) the loglikelihood, (2) the reduced chi-squared statistic ($\chi_{\bedit{r}}^2$),  (3) Bayesian information criterion \citep[BIC;][]{10.1214/aos/1176344136}, (4) and the Akaike information criterion \citep[AIC;][]{1100705}. Each of these metrics have different advantages and emphasize different features of the fitted models. Generally, each metric penalizes the number of free parameters in a different way and the AIC tends to be the least punitive of more complex models. \beditr{In our analysis, these different metrics all provide consistent results. Each metric is described in additional detail in the appendix.}

\subsection{Joint fit of astrometry and Radial Velocities}\label{joint_astrometry_rv_fit}
In order to further investigate the possibility of a long-period companion to HAT-P-2 b, we performed a joint fit of the RV and astrometric data sets. The model used to perform this joint analysis is described in detail in \citet{2021Venner}. In summary, we jointly fit the two datasets with a \bedit{three}-body Keplerian model \bedit{with 22 variable} parameters (the stellar mass and parallax, \bedit{which were assigned Gaussian priors of $1.33\pm0.03~M_\odot$ and $7.781\pm0.012$~mas respectively}; $P$, $K$, $e$, $\omega$, $t_p$ for both companions; orbital inclination ($i$) and longitude of ascending node ($\Omega$) \bedit{for the outer companion;} proper motion of system barycentre parameters ($\mu_{\text{bary,RA}}$,  $\mu_{\text{bary,Dec}}$); and normalization and jitter parameters for each RV dataset). \bedit{In comparison to the models described in Section \ref{pipeline_descript}, this is equivalent to a stable two-planet Keplerian fit.}

%\begin{figure}[ht!]
%\centering
%	\includegraphics[width=1.5in]{Figures.png}
%    \caption{Plot showing the archi^2tecture of our orbital %fitting pipeline}
%    \label{fig:cnnarchi{bedtecture}
%\end{figure}

%\begin{sidewaystable}
%\begin{sideways}
\begin{table*}[ht!]
   \centering
\begin{tabular}{l|lll}
Model& (i) Stable & (ii) Stable  &  (iii) Stable  \\
& 1 Planet  & 2 Planets & 2 Planets  \\
& (Keplerian)  & (2 Keplerians) & (Keplerian + Quadratic) \\
\hline
\hline
Goodness of fit metrics: &    &   &      \\
Reduced $\chi_{\bedit{r}}^2$ & 2.806 & 1.035      &     1.045 \\
loglikelihood  & -518.77      & -432.05  &  -432.87 \\
$\Delta$BIC$^\dagger$   & 165.18      & 14.90    &  21.16     \\
%$BIC  & 1074.631      & 924.349    &  930.617     \\
$\Delta$AIC$^\dagger$   & 175.72     & 12.26    &  15.90  \\
%AIC  & 1053.553      & 890.098    &   893.731   \\

\hline
MCMC Values: &      &   &     \\
\bedit{Mp ($M_j$)}           & $9.233_{-0.098}^{+0.099}$           & $9.033_{-0.099}^{+0.010}$       &    $8.998_{-0.098}^{+0.010}$           \\
%LogP    & $0.750775956_{-4.9e-08}^{+4.9e-08}$   & $0.750775955_{-4.9e-08}^{+4.9e-08}$               & $0.750775961_{-4.9e-08}^{+4.9e-08}$  & $0.750775956_{-4.9e-08}^{+4.9e-08}$  \\
P (days)      & $5.63346961_{-0.00000064}^{+0.00000064}$    & $5.63346961_{-0.00000064}^{+0.00000064}$     &  $5.63346960_{-0.00000064}^{+0.00000064}$       \\
Tp -2455000  (days)     & $289.46980_{-0.00034}^{+0.00034}$     & $289.48642_{-0.00033}^{+0.00033}$     &  $289.48642_{--0.00034}^{+0.00034}$  \\
e0      & $0.4990_{-0.0086}^{+0.0085}$        & $0.5003_{-0.0089}^{+0.0090}$  &   $0.4960_{--0.0088}^{+0.0088}$        \\
%de/dt$^*$ (1/day)  &              &              &        \\
%$de^2/dt^2$   (1/day$^2$)          &              &                  &     & $2.98e-09_{-2.7e-09}^{+2.6e-09}$       \\
$\omega$0  ($^{\circ}$)   & $188.25_{-0.39}^{+0.39}$         & $189.62_{-0.40}^{+0.40}$      &   $189.49_{-0.41}^{+0.41}$        \\
%d$\omega$/dt$^*$   ($^{\circ}$/day)             &              &            &                 \\
%$dom^2/dt^2$   ($^{\circ2}$/day$^2$)          &              &                  &     & $-2.5e-07_{-2.1e-07}^{+2.1e-07}$       \\
$\gamma_{\text{HARPS-2013-2015}}$      & $-26.8_{-8.6}^{+8.5}$         & $51_{-17}^{+19}$       &    $150_{-16}^{+16}$    \\
$\gamma_{\text{HARPS-2015-2017}}$      & $-20.8_{-18}^{+18}$         & $35_{-23}^{+24}$       &  $136_{-22}^{+22}$      \\
$\gamma_{\text{HIRES}}$      & $-33.8_{-4.9}^{+4.9}$           & $-170_{-14}^{+17}$       &   $77.4_{-9.8}^{+9.9}$    \\
linear \bedit{($\dot{\gamma}$)}     &             &               &    $-0.117_{-0.011}{+0.011}$         \\
quadratic \bedit{($\ddot{\gamma}$)}   &             &               &   $0.0000194_{-0.0000023}^{+0.0000023}$       \\
\bedit{M$_{p2}$ ($M_j$)} &    & $12.1_{-2.2}^{+2.9}$      &    \\
%LogP2  &         & $3.88_{-7.0e-02}^{+5.4e-02}$    &         \\
P2  (days)    &             & $8721_{-1141}^{+1407}$  &  \\
Tp2  -2455000  (days)    &             & $-9092_{-400}^{+287}$     &       \\
e02 &        & $0.313_{-0.069}^{+0.080}$ &  \\

$\omega$02 ($^{\circ}$) &          & $44_{-18}^{+17}$     &       \\
\end{tabular}
    \caption{Converged MCMC Results for Each of the Stable Orbital Models and their Goodness-of-!t Values. For each of the models, we specify whether the planets were modeled as a Keplerian or a quadratic trend.\\
    \textdagger \footnotesize{ \bedit{The $\Delta$BIC and $\Delta$AIC are computed by subtracting the BIC, AIC respectively corresponding to model (iv) 2 Planets (Kepelerian + Quadratic)  from the BIC, AIC values. The results for model (iv) are listed in Table \ref{Results_evolv_table}.} }}
    %* \footnotesize{Only the orbital parameters of HAT-P-2 b are allowed to vary over time.}}%\\
    %\textdagger \footnotesize{ This is the table where we use Gaussian Priors on P and T$_p$ based on the spitzer transits.}}
    \label{Results_table}
\end{table*}
%\end{sideways}

\begin{table*}[ht!]
   \centering
\begin{tabular}{l|lll}
Model & (iv) Evolving & (v) Evolving  &  (vi) Evolving  \\
&  1 Planet & 2 Planets & 2 Planets\\
 & (Keplerian) & (2 Keplerians) & (Keplerian + Quadratic) \\
\hline
\hline
Goodness of fit metrics:     &      \\
Reduced $\chi_{\bedit{r}}^2$ &  2.504 &   0.963  & 0.891 \\
loglikelihood  & -502.788 & -429.691  & -426.918   \\
$\Delta$BIC$^\dagger$    &  142.47 & 10.181  & 0.00    \\
%BIC$^\dagger$    &  1051.923 & 919.634  & 909.453    \\
$\Delta$AIC$^\dagger$    & 147.74 & 7.546  & 0.00  \\
%AIC$^\dagger$    & 1025.576 & 885.382  & 877.836  \\
\hline
MCMC Values: &  &   &   \\
%\bedit{Semi-amplitude} (\ms)                & $934_{-1.4e+01}^{+1.5e+01}$    &  & $913.4_{-1.4e+01}^{+1.4e+01}$      \\
Mp ($M_j$)    & $9.27_{-0.10}^{+0.10}$   & $9.02_{--0.10}^{+0.10}$ & $9.04_{-0.10}^{+0.10}$      \\
%LogP    & $0.750775956_{-4.9e-08}^{+4.9e-08}$   & $0.750775955_{-4.9e-08}^{+4.9e-08}$               & $0.750775961_{-4.9e-08}^{+4.9e-08}$  & $0.750775956_{-4.9e-08}^{+4.9e-08}$  \\
P (days)    & $5.63365_{-0.000040}^{+0.000039}	$ & $5.63346964_{-0.00000062}^{+0.00000063}$ & $5.633587_{-0.000043}^{+0.000044}$     \\
Tp  (days)    & $289.440_{-0.012}^{+0.011}$ & $289.48643_{-0.00034}^{+0.00033}$ & $289.48642_{--0.00034}^{+0.00034}$   \\
e0         & $0.499_{-0.011}^{+0.010}$      &  $0.502_{-0.012}^{+0.011}$ & $0.495_{-0.011}^{+0.011}$    \\
% 1/year unit version
de/dt$^*$ (1\textcolor{black}{/year})   & \textcolor{black}{\dedtNoPlanetC} & \textcolor{black}{\dedtKepPlanetC}  & \textcolor{black}{\dedtQuadPlanetC}   \\
% comment out the 1/day unit version
%de/dt$^*$ (1/day)   & $4.8_{-4.6}^{+4.7} \textcolor{black}{\cdot 10^{-6}}$ & $-2.0_{-5.0}^{+5.1} \textcolor{black}{\cdot 10^{-6}}$  & $2.1_{-5.1}^{+5.1} \textcolor{black}{\cdot 10^{-6}}$   \\
%$de^2/dt^2$   (1/day$^2$)          &              &                  &     & $2.98e-09_{-2.7e-09}^{+2.6e-09}$       \\
$\omega$0  ($^{\circ}$)   & $184.6_{-1.1}^{+1.1}$ &   $188.80_{-0.58}^{+0.57}$ & $185.9_{-1.2}^{+1.2}$       \\
% 1/year unit version
d$\omega$/dt$^*$   ($^{\circ}$\textcolor{black}{/year})     & \domdtNoPlanetC & \textcolor{black}{\domdtKepPlanetC} & \textcolor{black}{\domdtQuadPlanetC}       \\
% comment out the 1/day unit version
%d$\omega$/dt$^*$   ($^{\circ}$/day)     & $3.8_{-0.59}^{+0.59}\textcolor{black}{\cdot 10^{-3}}$ & $6.27_{-3.0}^{+3.0} \textcolor{black}{\cdot 10^{-4}}$  & $2.09_{-0.65}^{+0.65} \textcolor{black}{\cdot 10^{-3}}$       \\
%$dom^2/dt^2$   ($^{\circ2}$/day$^2$)          &              &                  &     & $-2.5e-07_{-2.1e-07}^{+2.1e-07}$       \\
$\gamma_{\text{HARPS-2013-2015}}$     & $-23.1_{-8.7}^{+8.5}	$ & $45_{-16}^{+18}$  & $142_{-16}^{+16}$       \\
$\gamma_{\text{HARPS-2015-2017}}$      & $-24_{-19}^{+19}$ &  $41.9_{-22}^{+23}$  & $134_{-23}^{+23}$      \\
$\gamma_{\text{HIRES}}$    & $-35.9_{-5.0}^{+5.0}	$ &  $-20.8_{-13}^{+15}$    & $71_{-10}^{+10}$      \\
linear \bedit{($\dot{\gamma}$)}     & & & $-0.110_{-0.011}^{+0.011}$    \\
quadratic \bedit{($\ddot{\gamma}$)}    &   &  & $1.78_{-0.25}^{+0.24} \textcolor{black}{\cdot 10^{-5}}$    \\
M$_{p2}$ ($M_j$)   &  & $11.6_{-2.0}^{+2.8}$ &  \\
%LogP2     &  &  &  \\
P2  (days)  &  & $8777_{-1200}^{+1400}$ &  \\
Tp2  (days)  &  & $-900_{-380}^{+300}$ &  \\
e02   &  & $0.346_{-0.072}^{+0.078}$ &  \\
$\omega$02 ($^{\circ}$)   &  & $43.0_{-19.5}^{+16.0}$ &  \\
\end{tabular}
    \caption{Converged MCMC Results for Each of the Evolving Orbital Models and their Goodness-of-!t Values. For each of the models, we specify whether the planets were modeled as a Keplerians or a quadratic trend. \\ 
    * \footnotesize{Only the orbital parameters of HAT-P-2 b are allowed to vary over time.} \\
    \textdagger \footnotesize{ \bedit{The $\Delta$BIC and $\Delta$AIC are computed by subtracting the BIC, AIC respectively corresponding to model (iv) 2 Planets (Kepelerian + Quadratic) from the BIC, AIC values.}}}%\\
    %\textdagger \footnotesize{ This is the table where we use Gaussian Priors on P and T$_p$ based on the spitzer transits.}}
    \label{Results_evolv_table}
\end{table*}

Cursory examination of the model reveals a degeneracy between $P$ and $e$ for the outer companion, arising due to incomplete coverage of the orbit. \beditr{If the outer planet has a shorter orbital period, it is more likely to be observed during its periastron passage than for longer orbital periods. We account for this in the same way as \citet{2019Blunt} and \citet{Venner2022} by adopting an informed prior on the orbital period such that:}

\begin{equation}
    \textcolor{black}{\mathcal{P}(P, t_d, B)} = \begin{cases}
    \MakeBox{\textcolor{black}{1}} & \textcolor{black}{\text{if}  (P-t_d) < B}\\
    \textcolor{black}{(B + t_d)/P} & \textcolor{black}{\text{otherwise}}
    \end{cases} 
\end{equation}
\begin{comment}
\begin{align}
    \textcolor{black}{\mathcal{P}(P, t_d, B)} &= \begin{cases}
    \MakeBox{\textcolor{black}{1}} & \textcolor{black}{\text{if}  (P-t_d) < B}\\
    \textcolor{black}{(B + t_d)/P} & \textcolor{black}{\text{otherwise}}
    \end{cases}
\end{align}  
\end{comment}

\beditr{where $\mathcal{P}$ is the probability of observing periastron passage, $P$ is the orbital period, $t_d$ is the duration of periastron passage, and $B$ is the baseline of observations. This prior is analogous to the period priors used for long-period transiting exoplanets \citep[e.g.][]{Vanderburg2016, Kipping2018}. Imposing this prior slightly penalizes orbital solutions where} the periastron passage of a long orbit happens to occur during the comparatively short span of observations. This discourages our model from fine-tuned long-period orbital solutions for the outer companion, without excluding them entirely. \beditr{\citet{2019Blunt} tested a range of values for $t_d$ and found that their orbital fits were indistinguishable, and thus chose $t_d=0$. We do the same in this analysis.} 

%To penalise improbable orbital configurations in which the periastron passage of a long orbit happens to occur during the comparatively short span of observations, we adopt an informed period prior used by \citet{2019Blunt} and \citet{Venner2022}.

We \bedit{again use \texttt{edmcmc} to explore the parameter space of the model}. We produce posterior samples by discarding the first 25\% of the chain as burn-in and save every hundredth step for each walker. We extract the median and 68.3\% confidence intervals for the parameters from the samples and compute the same confidence intervals for physical parameters
that can be derived (i.e. companion mass).

\section{Results}\label{Results}

\begin{comment}
\begin{figure*}[ht!]
\centering
	\includegraphics[width=0.9\textwidth]{figures/Stable, 1 Planet Model, 3 Instrument(s) 22_33_17PM.png}
    \caption{Plot of stable orbit model, one planet. On the left-hand panel, we plot the orbital fit of HAT-P-2 b and the residuals of the data minus the median model. In the right-hand panel, we plot linear and quadratic background trend fit and the RVs minus HAT-P-2 b. For both panels, the median model is plotted in dark purple and a 100 draws from the posteriors are plotted in light purple.}
    \label{fig:stable_1P}
\end{figure*}

\begin{figure*}[ht!]
\centering
	\includegraphics[width=0.9\textwidth]{figures/Stable, 2 Planet Model, 3 Instrument(s) 22_26_00PM.png}
    \caption{Plot of stable orbit model, two planets. On the left-hand panel, we plot the orbital fit of HAT-P-2 b and the residuals of the data minus the median model. In the right-hand panel, we plot the orbital fit of a potential outer companion and the RVs minus HAT-P-2 b. For both panels, the median model is plotted in dark purple and a 100 draws from the posteriors are plotted in light purple.}
    \label{fig:stable_2P}
\end{figure*}
\end{comment}

\subsection{Comparing Radial Velocity Orbital Fits}
For each of our four models, we computed the performance metrics described in Section \ref{perform_met} and listed them in Tables \ref{Results_table} and \ref{Results_evolv_table}. First, we can compare the stable, one-planet model with the stable model that allows for a long-period companion \bedit{modeled as either a Keplerian or a quadratic trend}. We find that the stable, two-planet model \bedit{(2 Keplerians)} has a significantly lower reduced $\chi_{\bedit{r}}^2 = \bedit{1.035}$, lower \bedit{$\Delta$BIC$=14.90$}, lower \bedit{$\Delta$AIC$=12.26$}, and higher loglikelihood$=\bedit{-432.05}$ compared to the one-planet model ($\chi_{\bedit{r}}^2 = \bedit{2.806}$, \bedit{$\Delta$BIC$=165.18$}, \bedit{$\Delta$AIC$=175.72$}, loglikelihood$=-518.7$). Furthermore, the stable, two-planet model \bedit{(Keplerian + Quadratic)} has a similar reduced $\chi_{\bedit{r}}^2 = \bedit{1.045}$, similar \bedit{$\Delta$BIC$=21.16$}, \bedit{$\Delta$AIC$=15.90$}, and  similar loglikelihood$=\bedit{-432.87}$ compared to the other two-planet model. To visually examine these orbital models, we plot the median fit along with 100 random draws for \bedit{all three} of these models in Figure \ref{fig:all_models}\bedit{a-f}. Overall, \bedit{the} two-planet models (either the 2 Keplerian or Keplerian + Quadratic version) are favoured in the comparison of these stable orbit models.

Next, we can compare various evolving orbit scenarios for HAT-P-2 b. We include the one-planet \bedit{(Keplerian)}, two-planet \bedit{(2 Keplerians), and two-planet (Keplerian + Quadratic)} scenario where the \bedit{HAT-P-2b}'s $e$ and $\omega$ can vary \bedit{linearly} with time. We find that the evolving, one-planet model ($\chi_{\bedit{r}}^2 = \bedit{2.504}$, \bedit{$\Delta$}BIC$=142.47$, \bedit{$\Delta$}AIC$=147.74$, loglikelihood $=-502.788$) provides a marginally better fit to the observations than a stable one-planet model. This is not a significant difference however. The evolving one-planet model is plotted in \bedit{Figure} \ref{fig:all_evolve_models}\bedit{a, b}. \bedit{Next, we find that model (v), the evolving 2 planets (2 Keplerians) model provides a slightly better fit ($\chi_{\bedit{r}}^2 = \bedit{0.963}$, \bedit{$\Delta$}BIC$=10.181$, \bedit{$\Delta$}AIC$=7.546$, loglikelihood $=-429.691$) than model (ii), the stable 2 planets (2 Keplerians) model.} Overall, we find that \bedit{model (vi), the} evolving, two-planet model \bedit{(Keplerian + Quadratic)} provides the best fit with the lowest $\chi_{\bedit{r}}^2 = \bedit{0.891}$, the lowest BIC$=\bedit{909.453}$, the lowest AIC$=877.836$, and the highest loglikelihood$=\bedit{-426.918}$. We plot this model in Figure \ref{fig:all_evolve_models}\bedit{e,f}. \bedit{We note that we also performed some additional tests to see how the rate of evolution for $e$, $\omega$ and their significance are impacted by how the long-term companion is modeled and include this in the discussion (Section \ref{model_choices_hatp2c})}.  %We note that we also investigated a second order evolving, two planet model, and found that this did not a better fit (reduced $\chi^2 = 0.992$, BIC$=927.6$, AIC$=882.891$) to the observations than a first order evolving, two planet model.

\begin{comment}
\begin{figure*}[ht!]
\centering
	\includegraphics[width=0.9\textwidth]{figures/2nd Order Evolving 1 Planet Model, 3 Instruments 19_52_12PM.png}
    \caption{Plot of 2nd order evolving orbit model (d$\omega$/dt, de/dt, d$\omega^2$/dt$^2$, de$^2$/dt$^2$). On the left-hand panel, we plot the orbital fit of HAT-P-2 b and the residuals of the data minus the median model. In the right-hand panel, we plot linear and quadratic background trend fit and the RVs minus HAT-P-2 b. For both panels, the median model is plotted in dark purple and a 100 draws from the posteriors are plotted in light purple.}
    \label{fig:evolving_2nd_ord}
\end{figure*}    
\end{comment}

\begin{figure*}[t!]
\centering
	\includegraphics[width=\textwidth]{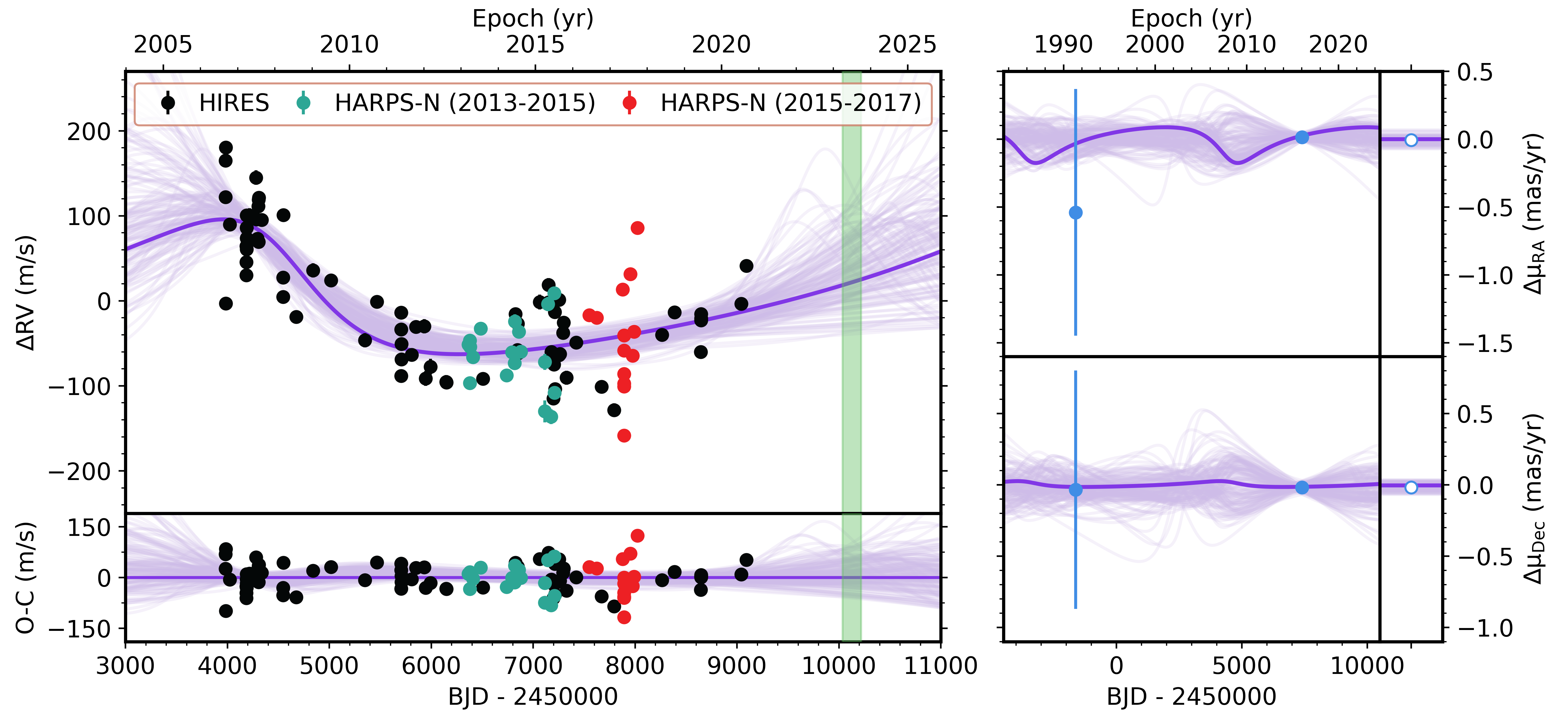}
    \caption{The orbit of HAT-P-2 c from a joint fit to the RVs (left) and astrometry (right). Colors are as in Figure \ref{fig:all_models}. The astrometry is consistent with zero net acceleration; the limits of this non-detection allow us to place useful constraints on the orbital period and inclination ($P_c=8500_{-1500}^{+2600}$ days, $i_c=90\pm16$ degrees). The resulting mass of HAT-P-2 c is $10.7_{-2.2}^{+5.2}$ $M_j$, establishing it as a \beditr{planetary-mass object}. We indicate the next observing window (April 1st, 2023 - September 28, 2023) with green shading. Further RV observations would help to improve constraints on the \beditr{companion's} orbit.
    \vspace{4mm}}
    \label{fig:astrometry_rv_fit}
\end{figure*}

\subsection{Astrometric-Radial Velocity joint fit}

\bedit{The results of the previous section strongly suggest the presence of an outer companion in the system, as has been found by past authors \citep{Lewis2013, Knutson2014, 2017Bonomo, Ment2018}. Furthermore, our 5-year extension of the HIRES timeseries allows us to clearly observe non-linearity in the RV acceleration. Previously, \citet{Knutson2014} constrained the mass of the outer companion to $8-200$ $M_j$ and the semi-major axis to $4-31$ AU based on the non-detection of a stellar companion in direct imaging. The strong orbital motion suggests that the companion has a relatively short period, and hence a mass on the lower end of this range.}

HAT-P-2 has been observed by both \textit{Hipparcos} and \textit{Gaia}, allowing us to make use of \textit{Hipparcos-Gaia} astrometry \citep{2018Brandt, Brandt2021} to place additional constraints on the outer companion. \beditr{We expect that even the lowest-mass solutions should produce a detectable ($\gtrsim$0.2~mas~yr$^{-1}$) astrometric signal based on the RV signal, but surprisingly we find that the \textit{Hipparcos-Gaia} astrometry of HAT-P-2 is consistent with zero net acceleration. A possible explanation for this is that the orbital period is close to the 25-year astrometric timespan. Specifically, if the orbital inclination is close to edge-on and the outer companion is at conjunction during both the \textit{Hipparcos} and \textit{Gaia} observations, then there will be no net acceleration between the \textit{Gaia} proper motion and \textit{Hipparcos-Gaia} mean proper motion, as observed.}

\begin{figure}
    \centering
    \includegraphics[width=\columnwidth]{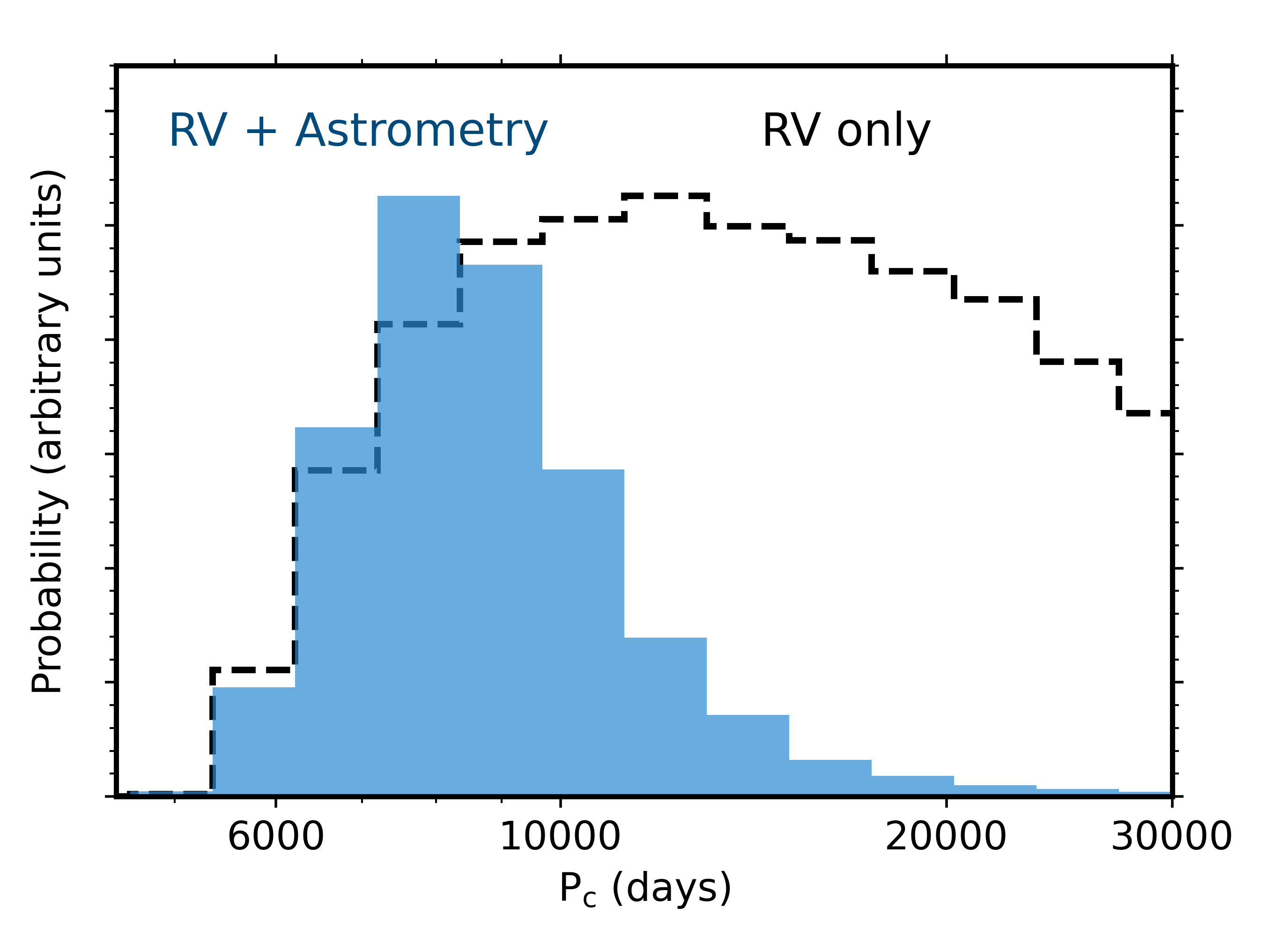}
    \caption{\bedit{The orbital period distribution of HAT-P-2 c for our joint RV + astrometry fit (blue) vs. RV-only (black).}}
    \label{fig:period_hist}
\end{figure}

This is borne out by our joint fit to the RVs and astrometry, the results of which we plot in Figure \ref{fig:astrometry_rv_fit}. \beditr{As anticipated, we find an orbital period of the outer companion of $23.2_{-4.1}^{+7.1}$ yr, comparable to 25 year timespan of the astrometric data. To demonstrate the importance of the astrometry for this result, in Figure \ref{fig:period_hist} we compare the period distribution from our joint fit with that of an RV-only model. The astrometric non-detection allows us to exclude orbital periods significantly longer than 15000 days at 95\% confidence. Furthermore, the non-detection constrains the orbital inclination to $\approx$90 degrees as this is the only orbital configuration consistent with zero net astrometric acceleration over the span of observations.}

The key posterior parameters of \beditr{the outer companion} are as follows:
\begin{equation}
\begin{split}
    &P_c = 8500_{-1500}^{+2600} \text{ days}\\
    &K_c = 89_{-13}^{+39} \text{ m/s} \\
    &e_c = 0.37_{-0.12}^{+0.13} \\
    &\omega_c = 38 \pm 32 \text{ degrees}\\
    &T_{p,c} = 2454150_{-800}^{+430} \text{ BJD}\\
    &i_c = 90\pm16  \text{ degrees.}
\end{split}
\end{equation}

%\begin{flalign}
%\begin{split}
%    &P_c = 8500_{-1500}^{+2600} \text{ days}\\
%    &K_c = 89_{-13}^{+39} \text{ m/s} \\
%    &e_c = 0.37_{-0.12}^{+0.13} \\
%    &\omega_c = 38 \pm 32 \text{ degrees}\\
%    &T_{p,c} = 2454150_{-800}^{+430} \text{ BJD}\\
%    &i_c = 90\pm16  \text{ degrees.}
%\end{split}
%\end{flalign}
%\linebreak
The longitude of node ($\Omega$) is unconstrained. \beditr{With the orbital period and inclination resolved, we find that the mass of this second companion} is $10.7_{-2.2}^{+5.2}$ $M_j$. The mass posterior is entirely substellar and corresponds to a \beditr{planetary-mass object, comparable in mass to HAT-P-2 b, which straddles the deuterium-burning limit at 13 $M_j$. We hence designate this companion as HAT-P-2 c, confirming the presence of a second planetary-mass companion in the HAT-P-2 system. We note that the parameters of HAT-P-2 c derived from this joint astrometric-RV fit agree within error with the RV fits in Tables \ref{Results_table}, \ref{Results_evolv_table}. The parameter values for HAT-P-2 c from the joint fit should be considered more robust.}

\section{Discussion}

\subsection{\bedit{Orbital Evolution}}
\subsubsection{Rates of orbital evolution}
\begin{figure}[ht!]
\centering
	\includegraphics[width=1\linewidth]{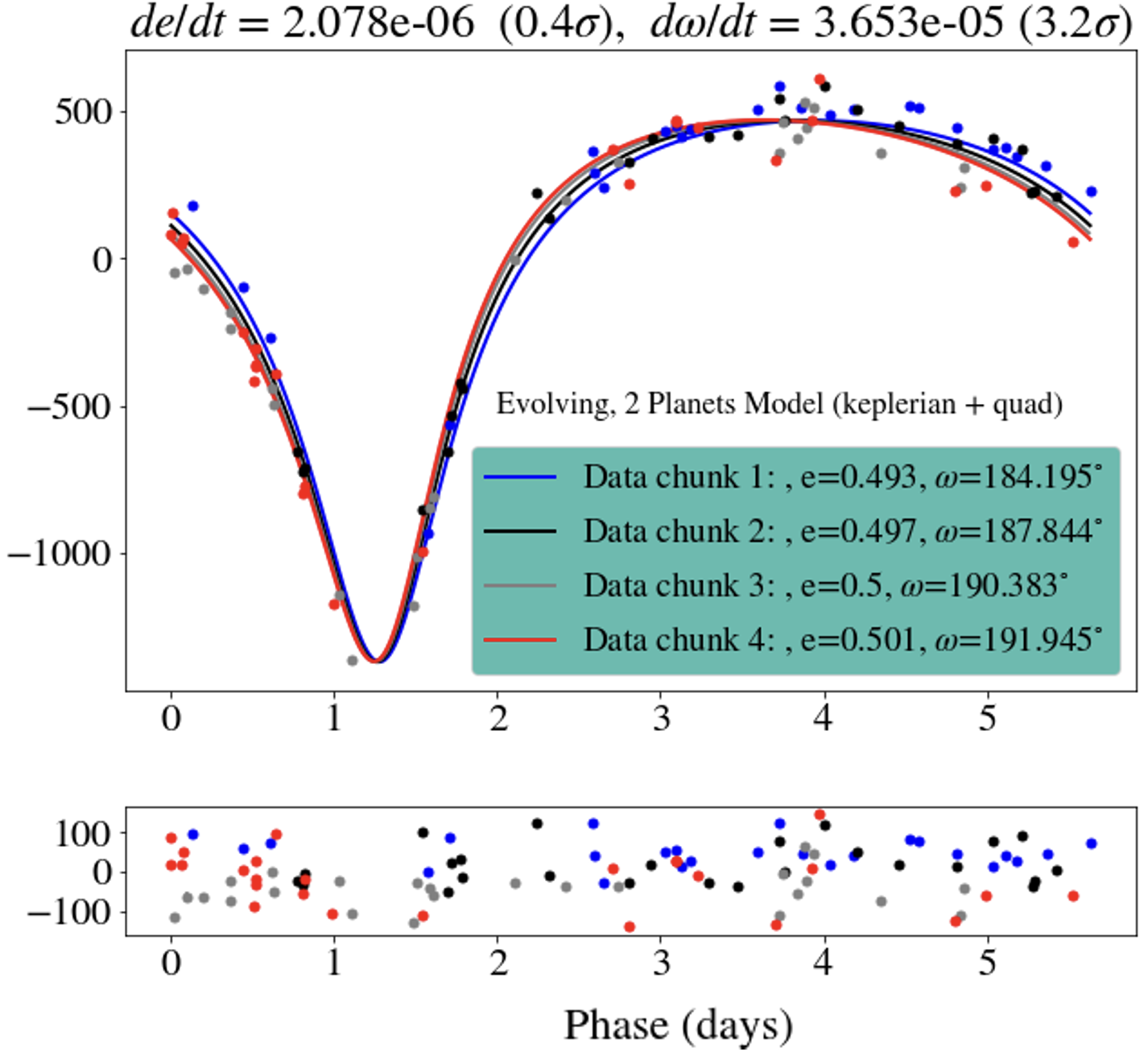}
    \caption{\bedit{Variation in orbital shape over time for HAT-P-2 as $e$ and $\omega$ increase linearly over time. The long-period companion is modeled as a quadratic trend here. The trends in $e$ and $\omega$ were fit using the entire dataset, but for visualization, the data is split into four equal-sized chunks ($\sim$ 25 observations per chunk) here. The median model corresponding to that time period is computed using the median $e$ and $\omega$ for the corresponding time period. }}
    \label{fig:money_plot}
\end{figure}

From our orbital fitting analyses, we find that although a two-planet\footnote{Where HAT-P-2 b is modeled as a Keplerian and HAT-P-2c is modeled as a quadratic trend}, evolving orbit model that allows $e$ and $\omega$ to vary linearly with time fits the RV measurements best\bedit{, the orbital parameters of HAT-P-2 c are poorly constrained in these RV models and follow-up observations are required to fully confirm or rule out orbital evolution in HAT-P-2 b}. \bedit{For this model, w}e find that $d\omega/dt = 
0.76_{-0.24}^{+0.24\circ}$/year and 
$de/dt = 7.67_{-18.5}^{+13.2} \textcolor{black}{\cdot 10^{-4}}$  /year which correspond to \bedit{3.23$\sigma$} and \bedit{$0.4\sigma$} detections respectively. \bedit{In Figure \ref{fig:money_plot}, we illustrate how these changes in $\omega$ and $e$ correspond to changes in the shape of the orbit over time.} \bedit{Although these rates of change agree within error with the rates ($d\omega/dt = 0.91 \pm 0.31$  $^{\circ}$/year, $de/dt = 8.9 \pm 2.8  \textcolor{black}{\cdot 10^{-4}}$/year) found in \citet{deWit2017}, they found a slightly more significant change in $e$}. Additional RV measurements in the next few years should allow for a further refinement of $de/dt$ and $d\omega/dt$ as described in \ref{sens_analysis}. \bedit{Specific choices in how HAT-P-2 c's long-term signal was modeled may also offer an explanation for the slight differences in evolution rates found in these different analyses as described in detail in Sections \beditr{\ref{deWit_compare} and} \ref{model_choices_hatp2c}.}%Figure \ref{fig:money_plot}, we illustrate how these changes in $\omega$ and $e$ correspond to changes in the shape of the orbit over time.

\begin{table*}[]
    \centering
    \begin{tabular}{l|cccc}
         \bedit{Model for HAT-P-2 c} & \bedit{None}   &\bedit{Linear} & \bedit{Quadratic} & \bedit{Keplerian} \\
         \hline
         \hline
         $\chi_r^2$ & 2.503  & 1.440 & 0.891 &  0.963\\ 
         $\Delta$BIC$^{\dagger}$  &  142.47 &  46.77 & 0.00 & 10.181\\ 
         %BIC$^{\textdagger}$  &  1051.923 &  956.223 & 909.453 & 919.634\\ 
         $\Delta$AIC$^{\dagger}$   & 147.74  &  49.405 & 0.00 & 7.5460363\\ 
         %AIC$^*$  & 1025.576  &  927.241 & 877.836 & 885.3820363\\ 
         \hline
         %de/dt (1/day) & $4.8_{-4.6}^{+4.7} \textcolor{black}{\cdot 10^{-6}}$  &  $9.0_{-4.8}^{+4.7} \textcolor{black}{\cdot 10^{-6}}$ & $2.1_{-5.1}^{+5.1} \textcolor{black}{\cdot 10^{-6}}$ & $-2.0_{-5.0}^{+5.1} \textcolor{black}{\cdot 10^{-6}}$\\
         
         de/dt (\textcolor{black}{1/year})   & \textcolor{black}{\dedtNoPlanetC} & \textcolor{black}{\dedtLinearPlanetC}  &  \textcolor{black}{\dedtQuadPlanetC}  & \textcolor{black}{\dedtKepPlanetC}  \\ 
         
         d$\omega$/dt ($^{\circ}$/\textcolor{black}{year})   & \textcolor{black}{\domdtNoPlanetC} & \textcolor{black}{\domdtLinearPlanetC} &  \textcolor{black}{\domdtQuadPlanetC}   & \textcolor{black}{\domdtKepPlanetC}  \\ 
         
         % comment out 1/day numbers
         %d$\omega$/dt ($^{\circ}$/day) & $3.1_{-5.9}^{+5.9} \textcolor{black}{\cdot 10^{-4}}$  &  $3.1_{-0.61}^{+0.60} \textcolor{black}{\cdot 10^{-3}}$ & $2.1_{-0.65}^{+0.65} \textcolor{black}{\cdot 10^{-3}}$ & $6.3_{-3.0}^{+3.0} \textcolor{black}{\cdot 10^{-4}}$\\ 
         $\sigma_{de/dt}$  &  1.03 &  1.90 & 0.41 & 0.32\\ 
         $\sigma_{d\omega/dt}$   & 5.20  &  5.11  & 3.23 & 2.11 \\ 
    \end{tabular}
    \caption{\bedit{Comparison of evolving orbit models where the second companion is not modeled or modeled as a linear trend, a quadratic trend, or a Keplerian.}\\
    \textdagger \footnotesize{ \bedit{The $\Delta$BIC and $\Delta$AIC are computed by subtracting the BIC, AIC respectively corresponding to model (iv) 2 Planets (Kepelerian + Quadratic) from the BIC, AIC values. Within this table, that is equivalent to subtracting the ``Quadratic'' model BIC, AIC values.}}}
    \label{tab:lin_quad_kep}
\end{table*}

\subsubsection{\beditr{Comparison to Radial Velocity analysis in \citet{deWit2017}}}\label{deWit_compare}
\beditr{In \citet{deWit2017}, the long-term trend corresponding the outer companion and the orbital evolution were modeled differently than in this analysis. In particular, the data available to \citet{deWit2017} only sampled the linearly decreasing component of the long-term trend and therefore \citet{deWit2017} modeled the outer companion as a linear trend rather than a quadratic or Keplerian. Now, with an additional five years of RV data, we were able to model the outer companion as a quadratic or Keplerian instead. For completeness, we also included a test where we approximate the outer companion as a linear trend as described in Section \ref{model_choices_hatp2c}.} 

\beditr{In addition to modeling the outer companion more accurately in our analysis, we also were able to model the orbital evolution in a more robust way with the availability of more RVs. \citet{deWit2017} checked for changes in $e$ and $\omega$ by i) splitting the RV data into two approximately equal halves (data prior and after BJD 2454604) and ii) fitting each dataset independently for an average value of $e$ and $omega$. \citet{deWit2017} then computed an estimate for the trends in $e$ and $omega$ based on the difference in values for each of these halves of the data. In this analysis, we instead fit the entire RV dataset simultaneously and define $e = de/dt \cdot time + e_0$, $\omega = d\omega/dt \cdot time + \omega_0$ such that we fit for $de/dt$ and $d\omega/dt$ directly. This approach should require that the trend is more persistent across the entire RV baseline and provide more robust estimates of the uncertainties on the trends as well.}

\begin{comment}
\begin{equation}
    de/dt = 
1.086e-03_{-1.371e-03}^{+1.317e-03}\text{/year}
\end{equation}

\begin{equation}
    d\omega/dt = 
0.798_{-0.229}^{+0.226} ^{\circ}\text{/year}
\end{equation}
\end{comment}

 %\footnote{\textcolor{black}{These values are not updated based on the most recent run listed in Table 1. Instead, they list the values from Table 3. With the new priors on P and T$_p$, the change in omega is only ~ 2$\sigma$ significance and I am a bit concerned. We should discuss.}}

\begin{figure*}[ht!]
\centering
\includegraphics[width=\textwidth]{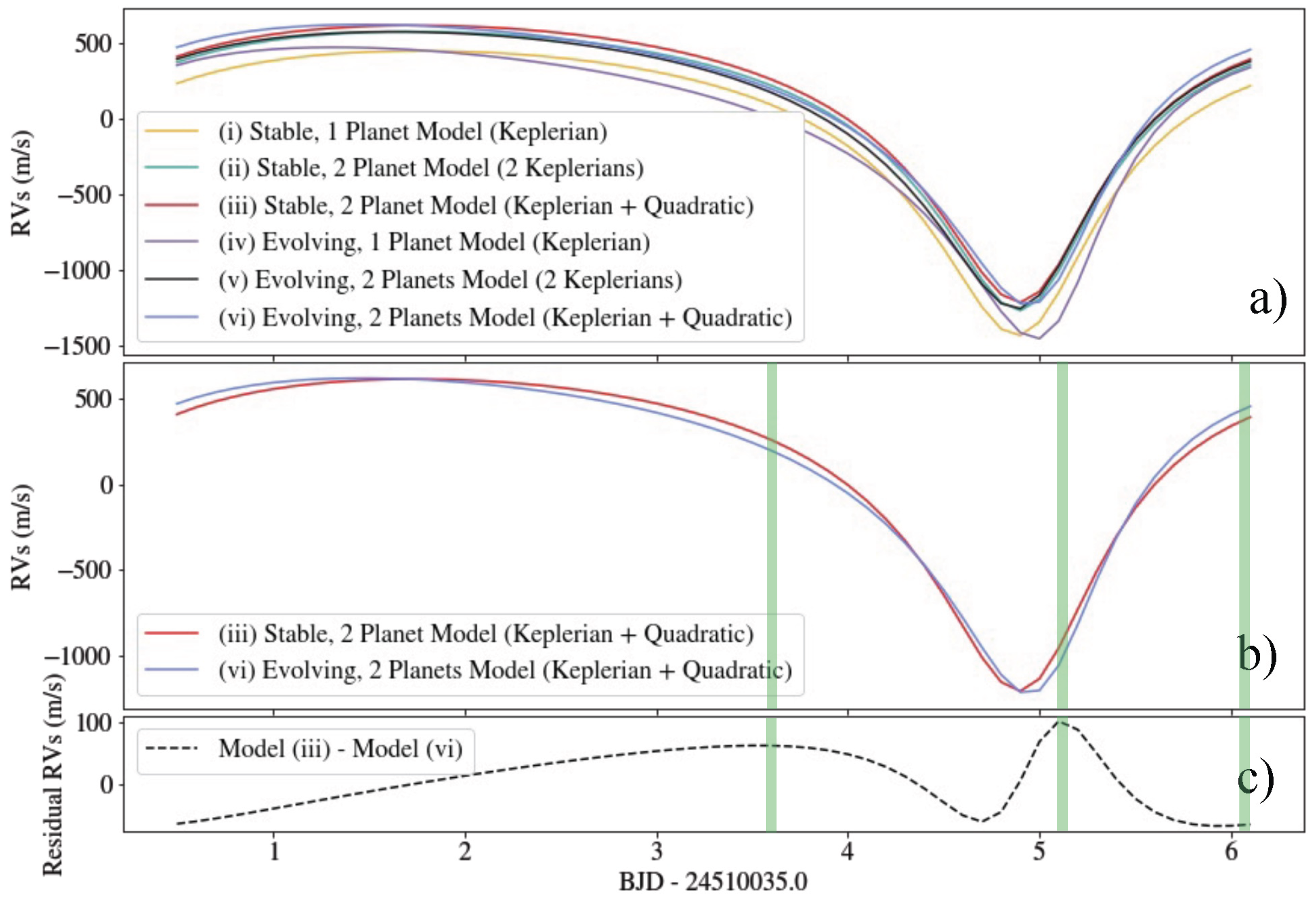}
    \caption{Expected RVs for the first orbit of the \beditr{current} observing \beditr{window} (April 1st, 2023 - April 5th, 2023). \bedit{(a)} This is a zoomed in version of Figure \ref{fig:propogated_models} and demonstrates the differences in expected RVs more clearly for a singular orbit. \bedit{(b) The two best-fit models (ii, vi) are plotted side by side to compare their differences. In green, we indicate the three orbital phases we propose to sample three times to detect the orbital evolution or lack thereof. (c) The residual RVs from subtracting the expected RVs from model (vi) from model (iii).}}
    \label{fig:sensi}
\end{figure*}

\begin{comment}
\subsection{Why does e seem to diverge significantly from literature values?}
In \cite{deWit2017}, the joint fit of HIRES RVs and Spitzer photometry provided $e = 0.51023 \pm 0.00042$. In \cite{2017Bonomo}, the analysis of HIRES and HARPS RVs provided $0.50833_{-0.00075}^{+0.00082}$ with priors on $P$ and $T_p$ from transits and occultations. When we include the HARPS-N RVs from \cite{2017Bonomo}, add additional HARPS-N RVs from the TNG Archive, and include all available HIRES RVs, we compute  $e = 0.957 \pm 0.009$ which is several sigma smaller than the literature values. 

\begin{figure}[ht!]
\centering
	\includegraphics[width=1\linewidth]{figures/e_explained_vertical.png}
    \caption{Eccentricity dependence on choice of RV datasets. In the top panel, only the HIRES RV Measurements are included and this leads to a higher median eccentricity than when both HIRES and HARPS-N RVs are included (bottom panel).}
    \label{fig:e_explained}
\end{figure}
\end{comment}

\subsubsection{Can the change in $\omega$ be explained by general relativity alone?}
We computed the rate of change expected from general relativity (GR) where
\begin{equation}
    \hat{\omega}_{GR} = \frac{3 G M_{*}}{a c^2 (1-e^2)} n
\end{equation}

\begin{equation}
    = 1.185e-2 \text{ degrees/year}
\end{equation}

where $n$ is the Keplerian mean motion, $a$ is the semi-major axis, and $M_*$ is the mass of the star. We find that for HAT-P-2 b, $\hat{\omega}_{GR} = 1.185e-2$ degrees/year. However, in our \bedit{best fit model}, we find  $\hat{\omega}= 0.76 \pm 0.24$ degrees/year, which is nearly \bedit{65} times the rate of change expected from GR. Thus, \bedit{this type of} rate of change cannot be explained by GR alone. \bedit{Tidal planet-star interactions may potentially be able to explain these types of} rapid change\bedit{s}. 

%according to \citet{2008Andres}

\begin{comment}
\subsection{Tidal planet-star interactions as a mechanism for rapid orbital changes}

In parallel to our RV analysis, we have been investigating the stellar evolution in HAT-P-2 using steller modeling codes \texttt{MESA}\citep{2011ApJS..192....3P} and \texttt{GYRE} \citep{2014IAUS..301..505T}. Bryan et al. (in prep) was able to reproduce the rates of $d\omega/dt$ and $de/dt$ with MESA by modeling the tidal interactions between the star and HAT-P-2 b. In addition, Bryan et al. (in prep) used GYRE to model the tidally excited oscillations in HAT-P-2 that were observed in \citep{deWit2017} and was able to reproduce the peaks at the 79th and 91st harmonics that were found in the secondary eclipse lightcurves from Spitzer \citep{deWit2017}. Both of these results point towards a tidal planet-star interaction origin of these rapid orbital changes.
\end{comment}

\subsubsection{\bedit{The impact of modeling choices for planet c on the rate of evolution of e and $\omega$}}\label{model_choices_hatp2c}
\bedit{Since previous analyses \citep{deWit2017, 2017Bonomo} modeled the long-period companion as linear or quadratic trends, we performed additional tests where we model the second companion in the same way and investigated its impact on the derived rate of orbital evolution. In Table \ref{tab:lin_quad_kep}, we list the derived rates of orbital evolution and their significance of detection when not modeling the outer companion and when modeling the outer companion as a linear trend, a quadratic trend, or as a Keplerian. We find that when we do not model HAT-P-2 c, the significance of the trend in $\omega$ (5.2$\sigma$) is highest and the significance of the trend in $e$ (1.03$\sigma$) is second-highest among these models. The derived rates of evolution are also among the highest amongst these models (de/dt = $1.75_{-1.68}^{+1.71}$ \textcolor{black}{$\cdot 10^{-3}$} / year, d$\omega$/dt = $1.12_{-0.22}^{+0.22}$ $^{\circ}$/year). When we model HAT-p-2 c as a linear trend, the significance of the trend in $e$ (1.9$\sigma$) increases and the significance of $d\omega/dt$ is marginally lower but still a 5.1$\sigma$ trend. The rates are the highest with de/dt = $3.28_{-1.75}^{+1.72}$ \textcolor{black}{$\cdot 10^{-3}$} / year and d$\omega$/dt = $1.12_{-0.22}^{+0.22}$ $^{\circ}$/year.} %Since the linear model yields the highest rates of evolution with some of the highest rates of significance, we visualize this evolution in Figure \ref{fig:money_plot}, which shows the orbital shape changes observed over time.}

\bedit{For the quadratic model, the significance of both trends decreases ($\sigma_{de/dt} = 0.41$, $\sigma_{d\omega/dt} = 3.23$) and the rates also  decrease (de/dt = $7.7_{-18.6}^{+18.6}$ \textcolor{black}{$\cdot 10^{-4}$} / year, d$\omega$/dt = $0.76_{-0.24}^{+0.24}$ $^{\circ}$/year). For the Keplerian model, we find that significance in $de/dt$ slightly increases ($\sigma_{de/dt} = 0.32$) and the significance in $d\omega/dt$ also decreases ($\sigma_{d\omega/dt} = 2.11$). The rate of evolution continues to decrease for $d\omega/dt$ (d$\omega$/dt = $0.228_{-0.11}^{+0.11}$ $^{\circ}$/ year) and direction of $de/dt$ changes (de/dt = $-7.3_{-18.0}^{+19.0}$ \textcolor{black}{$\cdot 10^{-4}$} / year). However, this trend in $e$ is so insignificant, indicating that that parameter is unconstrained by the data and so we do not attribute a physical interpretation to this change in sign.  Clearly, the exact modeling choices of HAT-P-2 c strongly affect the rates and significance of the evolution rates derived for HAT-P-2 b. The Keplerian model is the most physically motivated, but the period is still relatively unconstrained ($P_c = 8500_{-1500}^{+2600} \text{ days}$) and so are many of the other orbital parameters. Thus, further monitoring of HAT-P-2 system to allow us to model the long-term RV changes induced by HAT-P-2 c most accurately are necessary and should allow us to constrain the rates and possible evolution of HAT-P-2 b as well. In particular, we propose follow-up RVs to constrain the orbital parameters of HAT-P-2 c and follow-up transit and eclipse observations to solve precisely for the current values of $e$ and $\omega$, which when combined with archival Spitzer data should allow us to definitively constrain the rate of evolution.}

\subsubsection{Spitzer Transits and Secondary Eclipses}
To get additional constraints on $e$ and $\omega$ \bedit{and look for trends}, We re-processed and analyzed 4.5 $\mu$m Spitzer transit and secondary eclipse measurements using the pipeline described in \citet{Berardo2019}. We computed the best-fit systematic model using the pixel-level decorrelation (PLD) technique \citep{Deming2015}. We analyzed the primary and secondary eclipses by splitting the data into three chunks (2011, 2013, and 2015 observations) and \bedit{performed} three independent fits to \bedit{obtain} a measurements of $e$ and $\omega$ for each time period \bedit{and look for changes in their values}. For the 2011 observations, we find $e=0.5104 \pm 0.0025$ and $\omega = 189.10 \pm 2.19 ^{\circ}$. For 2015 observations, we find $e = 0.5128 \pm 0.0014$ and $\omega = 190.95 \pm 1.23 ^{\circ}$. Since the 2013 observations only included eclipses and no transits, we did not have the transit eclipse pair that is neccesary to constrain $e$ for the 2013 window. Overall the Spitzer transit and eclipse measurements provide some constraints on the possible rates of change for $e$ and $\omega$, but they only contain 3-4 transits/eclipses in each subset. With this little amount of data available for each of these subsets,  we do not get enough precision to accurately measure drifts in $e$. We plan to propose for JWST observations such that we can get the necessary precision to further constrain the rates of change of $e$ and $\omega$.

\subsubsection{RV Sensitivity Analysis}\label{sens_analysis}
In order to further characterize the HAT-P-2 system such that we can increase the significance of detection of the orbital evolution (or lack thereof), we performed a sensitivity analysis for the next observing window. In Figure \ref{fig:sensi}, we plot the expected RVs for the first orbit of the spring-fall 2023 observing season (April 1st, 2023 - April 5th, 2023). The expected RVs for a stable orbit will diverge significantly from an evolving orbit model and we should be able to clearly distinguish the two scenarios \bedit{as seen in Figure \ref{fig:sensi}b, c}. \bedit{In Figure \ref{fig:propogated_models}, we plot} the entire next observing window, \bedit{which} illustrate\bedit{s} that we should also be able to distinguish clearly between the one or two-planet models and \bedit{further constrain the period} of a potential HAT-P-2 c. \beditr{Thus, constraining the period of the long-period companion would require 3-4 RV measurements on HIRES/ Keck I. This could be done anytime in the observing window (April 1, 2023 - September 28, 2023) and would extend the RV baseline from 14 to 17 years. However, also detecting the orbital evolution would require additional RVs and would place constraints on the timing of the observations. In particular, to get a $6\sigma$ detection of the orbital evolution, we require nine new total RV measurements spanning three orbital periods (i.e. three RVs per orbit). The particular orbital phases are indicated in green in Figure \ref{fig:sensi}b,c. Although taking the RV observations at these orbital phases would be required, the observations could be taken for any orbit during the observing window.}

%\beditr{Further astrometric observations would be helpful if taken in a few years from now such that the outer companion is not at the expected time of conjunction. Then, we would likely detect an astrometric acceleration of the star by this long-period companion. This could provide additional constraints on its mass and period. Since we found from the analysis described in 5.1.4 that the derived rates and significance of orbital evolution are highly dependent on the modeling of the outer companion, astrometry measurements that constrain the outer companion’s parameters could indirectly also help constrain the orbital evolution.  }

% extend the baseline from 14 to 17 years

%More specifically, to get a $6\sigma$ detection of the orbital evolution, we require nine new total RV measurements spanning three orbital periods (i.e. three RVs per orbit). 

%subsubsection{Future Work and Prospects for these Methods on other Planetary Systems}
%We plan to incorporate transit observations into our orbital evolution fitting pipeline to allow users to perform simultaneous fits with transits and RVs. In addition, we plan to run our pipeline on other eccentric Hot Jupiter systems to look for signs of orbital evolution. Lastly, we hope to add additional features to our pipeline such as the ability to search for pulsations in lightcurves.

\subsection{\bedit{HAT-P-2 c}}
\bedit{The presence of a long-period outer companion to HAT-P-2 has been suggested in previous studies on the basis of a significant RV acceleration \citep{Lewis2013, Knutson2014, 2017Bonomo, Ment2018}. We have constrained the orbit of this companion for the first time by jointly fitting the RVs with astrometry from the \textit{Hipparcos-Gaia} Catalog of Accelerations \citep{2018Brandt, Brandt2021}. Though no acceleration is detected in the \textit{Hipparcos-Gaia} astrometry, this non-detection is sufficiently informative to constrain the orbit of the second companion. HAT-P-2 c is a $10.7_{-2.2}^{+5.2}$ $M_j$ \beditr{planetary mass object} with an orbital period of $P_c=8500_{-1500}^{+2600}$ days, placing it among the longest-period \beditr{substellar companions} that have ever been discovered from RVs}.

\bedit{The constraints of the astrometric non-detection requires HAT-P-2 c to have an edge-on orbital inclination ($i_c = 90\pm16$ degrees), as any other configuration would produce a detectable astrometric signal. This is comparable to the result of \citet{Errico2022}, who determined the true mass of the long-period planet HD~83443~c from an astrometric non-detection. HAT-P-2 c's orbital inclination is consistent with HAT-P-2 b, which is in turn has a low \beditr{projected} obliquity from the host star's rotational axis \citep{Winn2007b, Albrecht:2012}. This suggests that the orbital and rotational planes in the HAT-P-2 system are compatible with alignment, agreeing well with the conclusions of \citet{Becker2017}, who found that outer planets in Hot Jupiter systems must generally have low mutual inclinations to reproduce the low obliquities observed in these systems. The absence of strong misalignments in the HAT-P-2 system appears inconsistent with a chaotic dynamical history, which may be surprising considering the remarkably high orbital eccentricity of HAT-P-2 b. The formation history of this unusual planetary system will be an interesting topic to explore in future studies.}

\section{\bedit{Future Work}}

\bedit{We have proposed for RV follow-up observations to further constrain the period and orbital parameters of HAT-P-2 c, which will help us determine the magnitude and significance of orbital evolution for HAT-P-2 b (or lack thereof). In particular, we proposed to observe during specific phases of the orbit which have the information content necessary to disentangle between the one to two-planet and evolving or non-evolving scenarios as further described in Section \ref{sens_analysis}.}

\bedit{In addition, we plan to propose for a JWST partial phase curve of system with three goals in mind:}
\begin{enumerate}
    \item \bedit{Obtain ultra-precise transit and eclipse timing measurements: these measurements will enable us to get extremely tight constraints on the current values of $e$ and $\omega$, and thereby constrain their evolution.}
    \item \bedit{Measure the transient heating of HAT-P-2 b: Due to its high eccentricity and short period orbit, HAT-P-2 b is an ideal target for studying how atmospheres redistributes the time-varying heat flux from host stars. Analogous to \citet{Lewis2014}, we would propose measure the transient heating and use 3D atmospheric circulation models to further our understanding of the atmospheric response of exoplanets on highly eccentric orbits.} 
    \item \bedit{Follow-up on the planet-induced oscillations observed on the star noted in \citet{deWit2017} and check for wavelength-dependence of these oscillations. This will also allow us to construct a 2D heat map of HAT-P-2 b's surface, which was not possible in previous papers due to the complications added by the oscillations.}
\end{enumerate}

\section{Conclusion}
We developed an modular orbital evolution fitting pipeline that allowed us to perform apples-to-apples comparison between various dynamical scenarios that may explain observed radial velocity variation in the HAT-P-2 system. Using this pipeline, we confirmed that \bedit{the hints of rapid change} in argument of periastron ($\omega$) and eccentricity ($e$) \bedit{noted in \citet{deWit2017} appear to persist with an additional 5 years of RV data}. \bedit{However, the significance and exact rates of this evolution is highly dependent on the modeling of the long-period companion and requires additional follow-up observations to constrain. The largest} change\bedit{s} in $\omega$ are significantly larger than what would be expected solely from general relativistic precession and could be instead explained by tidal planet-star interactions. \bedit{Since these \beditr{precession} rates are not possible to explain with general relativity alone, further investigation to model the tidal planet-star interactions with stellar evolution and pulsation codes  (\citealt{2011ApJS..192....3P, 2014IAUS..301..505T},  \citet[\texttt{GYRE};][]{2014IAUS..301..505T}) is warranted}. We also ran a joint fit with \textit{Hipparcos-Gaia} \bedit{astrometry} and the RV measurements. This joint fit allowed us to put precise constraints on the outer companion in the system, which we find to be substellar: HAT-P-2 c. Furthermore, this work sets the stage for applying our orbital evolution fitting pipeline to a larger sample of stars that host eccentric Hot Jupiters, with the ultimate goal of modeling Hot Jupiter \beditr{precession} on a statistical scale. 

%We compared our observed orbital changes with steller evolution codes (\texttt{MESA, GYRE}) and were able to reproduce these rates of change in $\omega$ and $e$.

%% For this sample we use BibTeX plus aasjournals.bst to generate the
%% the bibliography. The sample631.bib file was populated from ADS. To
%% get the citations to show in the compiled file do the following:
%%
%% pdflatex sample631.tex
%% bibtext sample631
%% pdflatex sample631.tex
%% pdflatex sample631.tex

\section{Code availability}
The DE-MCMC Orbital evolution fitting pipeline developed for this manuscript is publicly available on Zenodo \citep{deBeurs_Zoe_github} and a living version is available on github\footnote{\hyperref[https://github.com/zdebeurs/Evolving_e_om_wobbling_edmcmc]{\url{https://github.com/zdebeurs/Evolving_e_om_wobbling_edmcmc}}}. The code was run on 7 cores of an M1 Macbook Air (16 GB RAM).

\section*{Acknowledgements}
\bedit{ZLD and JB would like to thank the generous support of the MIT Presidential Fellowship and to acknowledge that this material is based upon work supported by the National Science Foundation Graduate Research Fellowship under Grant No. 1745302. Work by JNW was supported by a NASA Keck PI Data Award.}

\bedit{This research has made use of the SIMBAD database and VizieR catalogue access tool, operated at CDS, Strasbourg, France. This research has made use of NASA's Astrophysics Data System. This work has made use of data from the European Space Agency (ESA) mission {\it Gaia} (\url{https://www.cosmos.esa.int/gaia}), processed by the {\it Gaia} Data Processing and Analysis Consortium (DPAC, \url{https://www.cosmos.esa.int/web/gaia/dpac/consortium}). Funding for the DPAC has been provided by national institutions, in particular the institutions participating in the {\it Gaia} Multilateral Agreement.}

\bibliography{Bibliography}{}
\bibliographystyle{aasjournal}

%\begin{comment}
\appendix

\begin{figure*}[ht!]
\centering
	\includegraphics[width=0.89\textwidth]{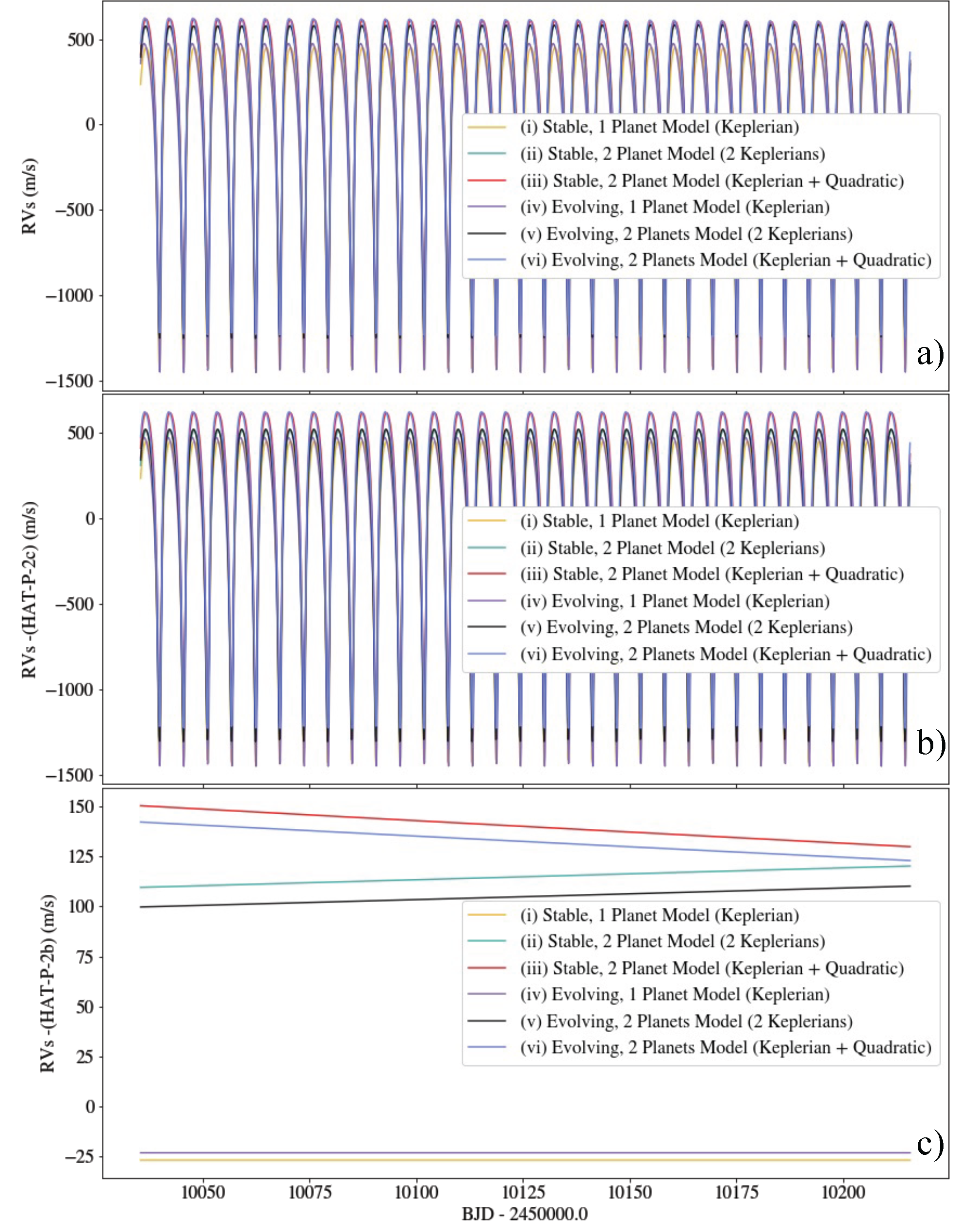}
    \caption{Expected RVs for each orbital model for the \beditr{current} observing \beditr{window} (April 1st, 2022 - September 28, 2023). (a) The expected RVs for the \bedit{6} orbital models \bedit{from Table \ref{Results_table}} are plotted for 32 orbits starting on April 1st, 2022. (b) The expected RVs minus the long-period companion are plotted. \bedit{c) The expected RVs minus HAT-P-2 c are plotted.}}
    \label{fig:propogated_models}
\end{figure*}

\beditr{In this appendix, we include the radial velocity measurements included in this analysis (Table \ref{RV_measurements}). In addition, we provide a supplemental figure of the expected RV measurements for each of the possible models during the next observing window (Figure \ref{fig:propogated_models}). Lastly, we include more detailed descriptions of the performance metrics used in evaluating each of our models (Section \ref{perform_metrics}).\\}

\section{\beditr{Performance Metrics}}\label{perform_metrics}
\beditr{We use the loglikelihood, reduced chi-squared statistic, Bayesian information criterion, and Akaike information criterion to evaluate the goodness of fit for our models. For completeness and pedagogical purposes, we include more detailed descriptions of each of these metrics here.}

\beditr{The loglikelihood ($\ln(L)$) is the natural log of the joint probability of the observations as a function of the parameters of the model. The likelihood is often written as $\L(\theta \ X)$ to emphasize that it is a function of the parameters $\theta$ given the data $X$. To find the most optimal $\theta$ given $X$, we sample a range of $\theta$ to maximize the likelihood. In practice, we do this by maximizing ($\ln(L)$) for practical purposes. In our case, the loglikelihood is defined as:}
\begin{equation}
    \textcolor{black}{\ln(L) = - \sum_i 0.5 * \frac{(x_i-\hat{x_i})^2}{\sigma^2} + \ln(\sigma)}
\end{equation}
\beditr{where $x_i$ is the $i$th observation in $x$ and $\sigma$ is the estimated variance.}

\beditr{The reduced chi-squared statistic ($\chi_{v}^2$) is one of the most widely used metrics to assess the goodness of fit. A lower $\chi_{v}^2$ indicates a better fit to the data.The $\chi_{v}^2$ is defined as:}
\begin{equation}
    \textcolor{black}{\chi_{v}^2 = \frac{1}{N-1} \sum_i \frac{(x_i - \hat{x_i})^2}{\sigma^2}}
\end{equation}
\beditr{where $N$ is the number of observations, $k$ is the number of free parameters, $x_i$ is the $i$th observation in $x$, $\hat{x_i}$ is the model prediction for the  $i$th observation, and $\sigma$ is the estimated variance. In general, a $\chi_{v}^2 > 1 $ corresponds to a model that does fully capture the complexity of the data and that error variance has been underestimated. In a model where $\chi_{v}^2 \ll 1 $, the model is overfitting and thus either (a) the error variance is overestimated or (b) the model is not properly fitting the noise in the observations. Generally, a $\chi_{v}^2$ close to $1$ corresponds to a proper fit and a lower value indicates a better fit to the data.} 
    
\beditr{The Bayesian information criterion \citep[BIC;][]{10.1214/aos/1176344136} is based on the likelihood function and adds a penalty term to the number of parameters in the models to reduce overfitting. Models with a lower BIC are generally preferred. The BIC is defined as }
\begin{equation}
    \textcolor{black}{BIC = k \ln (n) - 2 \ln(\hat{L})}
\end{equation}
\beditr{where $\hat{L}$ is the maximized value of the likelihood function, $n$ are the number of observations, and $k$ are the number of free parameters in the model.}

\beditr{The Akaike information criterion \citep[AIC;][]{1100705} also uses the likelihood function and is founded in information theory where AIC estimates the relative amount of information lost by a given model. The better model is the model where less information is lost. AIC allows one to strike a balance between optimizing the goodness of fit and the simplicity of the model. In this way, AIC allows one to deal with the problem of overfitting and underfitting the data. AIC is defined as}
\begin{equation}
    \textcolor{black}{AIC = k -  2 \ln(\hat{L})}
\end{equation}
\beditr{where the model with the lowest AIC value is the preferred model.}

\counterwithin{figure}{section}
\counterwithin{table}{section}

\counterwithin{figure}{section}

\begin{deluxetable}{cccc}[ht!]
\centering
\tablecaption{Summary of RV Measurements\label{harpsntable}}
%\begin{tabular}{llll}
%\tablehead{\multicolumn{4}{c}{ Summary of radial velocity measurements included in this analysis}}
\tablehead{\multicolumn{1}{c}{BJD-2450000} & \beditr{Instrument} & RV (m/s) & $\sigma_{RV}$ }
\startdata
%\hline
\hline
3981.77824  & HIRES      & -30.10    & 7.78        \\
3982.87244  & HIRES      & -324.11   & 7.00        \\
3983.81561  & HIRES      & 479.98    & 7.50        \\
3984.89572  & HIRES      & 826.89    & 7.31        \\
4023.69226  & HIRES      & 684.77    & 7.51        \\
4186.99899  & HIRES      & 648.04    & 7.82        \\
4187.1049   & HIRES      & 631.44    & 7.01        \\
4187.16062  & HIRES      & 654.71    & 7.01        \\
4188.01763  & HIRES      & 702.91    & 7.07        \\
4188.16036  & HIRES      & 721.16    & 6.89        \\
4189.01112  & HIRES      & 588.58    & 6.98        \\
4189.08965  & HIRES      & 595.95    & 6.90        \\
4189.15846  & HIRES      & 563.03    & 7.74        \\
4216.96014  & HIRES      & 660.49    & 8.42        \\
4279.87763  & HIRES      & 385.93    & 8.60        \\
4285.8246   & HIRES      & 109.99    & 5.58        \\
%4294.87944  & HIRES      & 719.76    & 5.38        \\
%4304.86572  & HIRES      & 572.48    & 6.12        \\
%4305.87086  & HIRES      & 712.56    & 5.79        \\
%4306.86596  & HIRES      & 714.55    & 6.73        \\
%4307.91312  & HIRES      & 431.68    & 6.23        \\
.  & .    & .   & .    \\
.  & .    & .   & .    \\
.  & .    & .   & .    \\
%7672.70624  & HIRES      & -795.90   & 7.42        \\
%7793.17997  & HIRES      & 518.55    & 8.04        \\
%7879.60946  & HARPS-N    & 331.82    & 1.31        \\
7891.38229  & HARPS-N    & -237.31   & 3.64        \\
7891.38885  & HARPS-N    & -127.17   & 3.62        \\
7891.39284  & HARPS-N    & -191.79   & 3.33        \\
7891.39647  & HARPS-N    & -180.93   & 3.36        \\
7891.68477  & HARPS-N    & -617.71   & 3.24        \\
7891.69311  & HARPS-N    & -594.11   & 1.86        \\
7953.48534  & HARPS-N    & -218.48   & 1.21        \\
7975.43735  & HARPS-N    & 221.07    & 1.43        \\
7989.35655  & HARPS-N    & 541.67    & 1.31        \\
8024.41654  & HARPS-N    & 775.37    & 1.20        \\
8263.12948  & HIRES      & -100.28   & 6.33        \\
8386.70729  & HIRES      & 208.80    & 6.09        \\
8645.77618  & HIRES      & 202.41    & 5.68        \\
8648.8643   & HIRES      & 586.11    & 6.54        \\
8648.86713  & HIRES      & 588.86    & 6.73        \\
8648.86985  & HIRES      & 594.18    & 6.21        \\
9042.84334  & HIRES      & 534.45    & 7.19        \\
9092.71825  & HIRES      & 58.63     & 7.35 \\
\enddata
\tablecomments{Full table available online.\label{RV_measurements}}
\end{deluxetable}

\end{document}